\pdfoutput=1
\documentclass[journal]{IEEEtran}
\usepackage[T1]{fontenc}
\usepackage{cite}
\usepackage{amsmath,amssymb,amsfonts}
\usepackage{algorithm}
\usepackage{algorithmic}
\usepackage{graphicx}
\usepackage{epsfig}
\usepackage{epstopdf}
\usepackage{textcomp}
\usepackage{xcolor}
\usepackage{subfigure}
\usepackage{booktabs}
\usepackage{multirow}
\usepackage{diagbox}
\usepackage{url}
\usepackage{threeparttable}
\usepackage{array}
\usepackage{stfloats}
\makeatletter
\def\UrlAlphabet{%
      \do\a\do\b\do\c\do\d\do\e\do\f\do\g\do\h\do\i\do\j%
      \do\k\do\l\do\m\do\n\do\o\do\p\do\q\do\r\do\s\do\t%
      \do\u\do\v\do\w\do\x\do\y\do\z\do\A\do\B\do\C\do\D%
      \do\E\do\F\do\G\do\H\do\I\do\J\do\K\do\L\do\M\do\N%
      \do\O\do\P\do\Q\do\R\do\S\do\T\do\U\do\V\do\W\do\X%
      \do\Y\do\Z}
\def\UrlDigits{\do\1\do\2\do\3\do\4\do\5\do\6\do\7\do\8\do\9\do\0}
\g@addto@macro{\UrlBreaks}{\UrlOrds}
\g@addto@macro{\UrlBreaks}{\UrlAlphabet}
\g@addto@macro{\UrlBreaks}{\UrlDigits}
\makeatother

\graphicspath{{figures/}}

\begin{document}
\setlength{\textfloatsep}{10pt}
\ifdefined \GramaCheck
  \newcommand{\CheckRmv}[1]{}
  \newcommand{\figref}[1]{Figure 1}%
  \newcommand{\tabref}[1]{Table 1}%
  \newcommand{\secref}[1]{Section 1}
  \newcommand{\algref}[1]{Algorithm 1}
  \renewcommand{\eqref}[1]{Equation 1}
\else
  \newcommand{\CheckRmv}[1]{#1}
  \newcommand{\figref}[1]{Fig.~\ref{#1}}%
  \newcommand{\tabref}[1]{Table~\ref{#1}}%
  \newcommand{\secref}[1]{Section~\ref{#1}}
  \newcommand{\algref}[1]{Algorithm~\ref{#1}}
  \renewcommand{\eqref}[1]{(\ref{#1})}
\fi
\newtheorem{theorem}{Theorem}
\newtheorem{proposition}{Proposition}
\newtheorem{assumption}{Assumption}
\newtheorem{definition}{Definition}
\newtheorem{condition}{Condition}
\newtheorem{property}{Property}
\newtheorem{remark}{Remark}
\newtheorem{lemma}{Lemma}
\newtheorem{corollary}{Corollary}

\title{Low-Overhead Receiver Design for Data-Dependent Superimposed Training via Deep Learning}

\author{Xinjie~Li,
        Xingyu~Zhou, \IEEEmembership{Graduate Student Member,~IEEE,}
        Jing~Zhang, \IEEEmembership{Member,~IEEE,}
        Chao-Kai~Wen, \IEEEmembership{Fellow,~IEEE,}
        Xiao~Li, \IEEEmembership{Member,~IEEE,}
        and Shi~Jin, \IEEEmembership{Fellow,~IEEE}
\thanks{X. Li, X. Zhou, J. Zhang, X. Li, and S. Jin are with the School of Information Science and Engineering, Southeast University, Nanjing 210096, China
(e-mail: lixinjie@seu.edu.cn; \protect \url{xy_zhou@seu.edu.cn}; jingzhang@seu.edu.cn; \protect \url{li_xiao@seu.edu.cn}; jinshi@seu.edu.cn).}
\thanks{C.-K. Wen is with Institute of Communications Engineering, National Sun Yat-sen University, Kaohsiung 80424, Taiwan
(e-mail: chaokai.wen@mail.nsysu.edu.tw).}
}

\maketitle

\begin{abstract}
Superimposed pilot (SIP) transmission improves spectral efficiency by eliminating the dedicated pilot overhead required in orthogonal pilot (OP)-based schemes. However, SIP suffers from severe pilot-data coupling, which leads to a critical performance-complexity bottleneck at the receiver. To address this issue, this paper proposes a low-overhead transmission framework that revitalizes data-dependent superimposed training (DDST) with enhanced interference mitigation strategies. First, for quasi-static block-fading channels, an enhanced DDST receiver is developed to achieve non-iterative pilot-data decoupling by exploiting data-dependent algebraic structures. Second, to overcome the sensitivity of conventional DDST to channel variations and symbol misidentification in fast time-varying environments, a mix transmission scheme is developed. By strategically applying DDST to a subset of resource elements, the proposed scheme combines the interference-free transmission property of OP with the zero-pilot-overhead advantage of SIP, thereby improving demapping reliability and interference suppression. Furthermore, under the proposed mix scheme, a Vision Transformer-based neural receiver is designed to capture the orthogonal structure between pilots and perturbation-bearing data, as well as the underlying channel correlations, thereby relaxing the stringent quasi-static assumption required for interference disentanglement. Simulation results demonstrate that the proposed framework achieves significant performance gains in the low-to-medium SNR regime under time-varying channels while providing superior computational efficiency compared with state-of-the-art SIP receivers.
\end{abstract}


\IEEEpeerreviewmaketitle

\section{Introduction}
 
\IEEEPARstart{C}{hannel} state information (CSI) acquisition is fundamental to the performance and spectral efficiency of multiple-input multiple-output orthogonal frequency division multiplexing (MIMO-OFDM) systems \cite{OPMar,GenZhou}. Conventional pilot-assisted schemes allocate orthogonal resource elements (REs) to pilot and data symbols \cite{OPTru,NRCite}, thereby avoiding mutual interference through orthogonal multiplexing. However, this orthogonal pilot (OP) paradigm incurs considerable pilot overhead, limiting spectral efficiency. 
To alleviate this limitation, superimposed pilot (SIP) transmission has emerged as a promising alternative by embedding pilots directly into data symbols in the power domain \cite{SIPXie}. Despite its high resource efficiency, the non-orthogonal nature of SIP inherently introduces severe pilot contamination and data interference, which substantially complicates receiver design.

Existing SIP receivers mainly address pilot-data coupling through two representative paradigms: iterative joint processing and data-driven deep learning (DL). On one hand, joint channel estimation, signal detection, and decoding (JCDD) frameworks employ iterative loops to improve estimation accuracy using refined data feedback \cite{SIPMa,SIPJing,SIPQian}. However, these JCDD-based approaches either emphasize theoretical optimality at the expense of practical implementation or suffer from performance degradation due to intrinsic design limitations. On the other hand, purely data-driven DL approaches learn latent signal structures directly from received observations to suppress pilot-data interference \cite{SIPAit,SIPXiao,GenZhou2}. Nevertheless, these heavily parameterized models generally exhibit limited scalability and degraded robustness in multi-stream and dynamically varying wireless environments. 
To bridge this gap, recent studies have integrated JCDD with DL to improve both feasibility and reliability \cite{SIPLi,SIPLi2}. In these approaches, a unified signal processing framework initializes the JCDD architecture using linear minimum mean square error (LMMSE)-based channel estimation together with soft-input soft-output detection and decoding, while a lightweight convolutional neural network (CNN) refines data-aided channel estimation. Although this hybrid design improves effectiveness and generalization capability, its iterative structure still incurs substantial computational complexity, hindering real-time deployment under stringent latency requirements.

Fundamentally, a trade-off between model complexity and inference latency remains inherent in SIP receiver design. Lightweight model-driven receivers generally rely on iterative processing to exploit feedback data priors for pilot refinement and interference suppression \cite{SIPLi,SIPLi2}. In contrast, existing non-iterative receivers typically require large neural networks to implicitly capture these priors from the non-Gaussian and nonlinear observations induced by signal superposition, often increasing training complexity and reducing reliability \cite{GenZhang}. Consequently, overcoming this performance-latency trade-off requires a new transmission and receiver design paradigm that simplifies interference disentanglement.

In this context, data-dependent superimposed training (DDST), a classical yet relatively underexplored transmission scheme, has recently regained attention as a promising solution to the above bottleneck. Unlike SIP schemes, which place the burden of interference suppression entirely on the receiver, DDST facilitates pilot-data disentanglement through transmitter-side structural design. In single-input single-output (SISO) systems under quasi-static channels, a data-dependent perturbation sequence is superimposed onto time-domain transmitted symbols to generate periodic frequency nulls for pilot insertion while completely canceling data energy \cite{DDSTGho2,DDSTWu}. As a result, pilot and data components can be decoupled at the receiver without iterative data feedback.
When extended to time-varying channels, however, Doppler-induced channel variations destroy the ideal frequency-null structure and spread the received training contribution into neighboring frequencies. To mitigate this effect, basis expansion models (BEMs) are commonly adopted to represent channel variations within a low-dimensional subspace, thereby confining the resulting leakage to limited frequency neighborhoods and enabling DDST perturbation design over the affected bins \cite{DDSTTvGho,DDSTTvHe,DDSTTvCar}. Although effective in alleviating the SIP dilemma, the introduced perturbation sequence inevitably distorts linear symbol recovery, causing severe detection degradation known as the symbol misidentification problem \cite{DDSTChan,DDSTDou,DDSTQing}.

Extending DDST to MIMO systems introduces further structural challenges. Directly adapting time-varying BEM-based DDST schemes to MIMO systems generally requires carefully separated frequency-null allocations across all transmit antennas. Meanwhile, the perturbation design must suppress both inter-carrier interference and multi-stream interference, leading to a highly constrained optimization problem that becomes increasingly difficult to solve analytically. Furthermore, existing hard-decision-based approaches for mitigating symbol misidentification \cite{DDSTChan,DDSTDou} exhibit limited scalability in high-order modulation and coded transmission scenarios. 
Consequently, practical DDST schemes for MIMO systems often enforce frequency-domain orthogonality between pilot and perturbation-embedded data \cite{DDSTGho,DDSTKam}. Specifically, by constructing the constraint matrix based on the pilot sequence, an orthogonal projection matrix can be derived as the perturbation operator. At the receiver, subspace projection is implicitly achieved through least-squares (LS)-based channel estimation, thereby decoupling pilot and data components. However, this decoupling critically relies on the quasi-static channel assumption. Under practical time-varying channels, the breakdown of the block-structured LS formulation prevents effective subspace projection, undermining the effectiveness of DDST.

Motivated by these algebraic challenges, DL techniques provide a promising direction for enhancing DDST reception through data-driven modeling rather than explicit analytical formulations. A preliminary proof of concept was presented in \cite{DDSTQing} for block-fading SISO systems, where deep neural networks were used to refine LS-based channel estimation and zero-forcing-based detection at the DDST receiver while compensating for nonlinear distortions induced by hardware impairments.
Building on this insight, advanced DL architectures are promising candidates for addressing the orthogonality collapse encountered in time-varying MIMO systems. In particular, the self-attention mechanism shows strong capability in modeling the latent nonlinear coupling structures induced by non-orthogonal signal superposition and in disentangling pilot and data components through learned structural priors. This potential was demonstrated in \cite{SIPZou}, where the proposed SIP receiver leveraged self-attention to capture long-range spatial correlations, enabling hierarchical feature extraction and effective multi-stream interference mitigation. These findings suggest a promising direction for addressing the inherent challenges in DDST-based transmissions.

Motivated by the above observations, this paper develops a low-overhead transmission framework by integrating DDST-based transmission with advanced DL techniques. To improve practicality in dynamic wireless environments, the proposed framework accommodates both block-fading and time-varying scenarios. For block-fading channels, the conventional DDST scheme is adopted, based on which a DL-enhanced receiver is developed to establish a reliable baseline. To address orthogonality collapse and symbol misidentification under time-varying conditions, a mix transmission scheme is further proposed by combining DDST-based and pure data REs with a self-attention-driven neural receiver. Positioned between the OP and SIP paradigms, the proposed framework exploits both the distortion-free transmission property of OP and the spectral efficiency advantage of SIP, enabling effective interference disentanglement.
The main contributions of this paper are summarized as follows:
\begin{itemize}
\item An enhanced DDST receiver is developed for coded MIMO-OFDM systems under block-fading channels. Specifically, an LMMSE-based channel estimation scheme is derived to improve estimation accuracy, while a hybrid CNN-long short-term memory (LSTM) detection network is designed to partially mitigate symbol misidentification. By exploiting the algebraic structure of DDST, the proposed receiver achieves pilot-data decoupling and establishes a feasible baseline for subsequent extensions.

\item A mix transmission scheme is proposed for time-varying channels, where REs are partitioned into DDST-based and pure data segments. The uncontaminated soft information extracted from pure data REs improves the reliability of extrinsic information and effectively alleviates symbol misidentification. This structural design compensates for the limitations of purely DL-based mitigation and improves demapping reliability during decoding.

\item Based on the proposed mix transmission scheme, a neural receiver is further developed. For channel estimation, a Vision Transformer (ViT)-based encoder and a CNN-based decoder are jointly designed to exploit structural features embedded in DDST REs for interference suppression and channel reconstruction. For symbol detection, two dedicated hybrid CNN-LSTM subnetworks are constructed to perform bit-level demapping over DDST and pure data REs independently.

\item Simulation results demonstrate that the proposed framework achieves reliable channel estimation and improved throughput. Furthermore, under appropriate resource allocation, the proposed mix scheme improves demapping reliability and reduces computational complexity compared with state-of-the-art SIP receivers.
\end{itemize}


\emph{Notations:} 
Superscripts ${( \cdot )^{\mathrm{T}}}$ and ${( \cdot )^{\mathrm{H}}}$ denote the transpose and conjugate transpose, respectively. $z^*$ and $| z |$ denote the complex conjugate and modulus of a complex number $z$. The expectation operator is denoted by $\mathbb{E}\{ \cdot \}$. $\mathbf{I}$ is the identity matrix, while $\mathbf{0}$ and $\mathbf{1}$ denote the all-zero and all-one matrices, respectively. 
$\mathbf{A}=[a_{i,j}]$ denotes a matrix $\mathbf{A}$ whose $(i,j)$-th element is $a_{i,j}$. In addition, $\otimes$ denotes the Kronecker product.

\section{DDST Receiver under Block-fading Scenarios}
\label{sec_block}
This section discusses the application of DDST to MIMO-OFDM systems. We first introduce the conventional DDST receiver in \cite{DDSTGho}, followed by the proposed enhanced DDST receiver, which improves both channel estimation and detection accuracy. 

\vspace{-1em}
\subsection{System Model}
\label{sec_block_system}
Consider a MIMO system with $N_\text{r}$ receive antennas and $N_\text{t}$ transmit antennas. An OFDM frame consisting of $K$ subcarriers and $T$ consecutive symbols is transmitted from each antenna. For the $n$-th transmit antenna, the training sequence of length $T$, denoted by ${{\mathbf{p}}_{n}} = [p_n(t)] \in \mathbb{C}^{T \times 1}$, is constructed, where the $t$-th element is given by ${p_n(t)} = e^{\frac{j2 \pi nt}{N_\text{t}}}$. The training symbols are subsequently superimposed onto the data sequences at each subcarrier to facilitate CSI acquisition \cite{DDSTGho}.

For data transmission, the information sequences are modulated using a complex $M$-ary quadrature amplitude modulation (QAM) constellation $\mathcal{A}$, where ${Q = \log_2 M}$ denotes the number of bits per complex symbol. The data vector corresponding to the $n$-th transmit antenna at the $k$-th subcarrier is denoted by ${{\mathbf{d}}_n^k} = [d_n^{k}( t )] \in \mathbb{C}^{T \times 1}$. It subsequently undergoes DDST processing to generate the transmitted data sequence ${{\mathbf{\widetilde{d}}}_n^k} = [{\widetilde{d}}_n^{k}( t )] \in \mathbb{C}^{T \times 1}$, given by
\begin{equation}
       {{\mathbf{\widetilde{d}}}_n^{k}} = {{\mathbf{d}}_n^{k}} - {{\mathbf{e}}_n^{k}} = \left(\mathbf{I}-{\mathbf{J}}\right) {{\mathbf{d}}_n^{k}}.
       \label{equ_dtilde_nk}
\end{equation} 
Here, the perturbation sequence ${{\mathbf{e}}_n^{k}} = [e_n^k( t )] \in \mathbb{C}^{T \times 1}$ is generated through the DDST processing matrix ${\mathbf{J}} \in \mathbb{C}^{T \times T}$ to suppress data interference on the training sequence \cite{DDSTKam}. The processing matrix is defined as\footnote{Notably, ${\mathbf{J}}$ can be equivalently expressed as ${{\mathbf{P}}^{\mathrm{H}}}{({\mathbf{P}}{{\mathbf{P}}^{\mathrm{H}}})^{ - 1}}{\mathbf{P}}$, such that ${\mathbf{I}}-{\mathbf{J}}$ corresponds to the orthogonal projection matrix onto the orthogonal complement of the row space of ${\mathbf{P}}$. This property intuitively explains the interference-suppression capability, as further detailed in \secref{sec_block_trad}.}
\begin{equation}
       {\mathbf{J}} = \frac{1}{P}{{\mathbf{1}}_P} \otimes {{\mathbf{I}}_{{N_{\text{cycle}}}}},
       \label{equ_J}
\end{equation}
where $P$ satisfies $P = {T/{N_{\text{cycle}}}}$, and ${N_{\text{cycle}}}$ denotes the cyclic period of the training pilot, i.e., ${N_{\text{cycle}}}={N_{\text{t}}}$. 
The transmitted symbol from the $n$-th antenna corresponding to the $(k,t)$-th RE is then expressed as 
\begin{equation} 
       s_n^k ( t ) = \sqrt \rho \, {p_n}( t ) + \alpha \, {\widetilde{d}}_n^{k}( t ),
       \label{equ_s_nkt}
\end{equation}
where $\rho \in (0,1)$ denotes the power allocation factor, and $\alpha$ is the data power normalization factor, expressed as
\begin{equation}
       \alpha = \sqrt {\frac{{1 - \rho }}{{1 - \frac{1}{P}}}} .
       \label{equ_alpha}
\end{equation}

After OFDM reception, signals from all $KT$ time-frequency REs are collected. Specifically, the received signal from the $m$-th antenna at the $k$-th subcarrier is denoted by ${\mathbf{y}}_m^k = [y_m^k ( t )] \in \mathbb{C}^{T \times 1}$, where the $t$-th element is expressed as
\begin{equation}
       y_m^k ( t ) = \sum\limits_{n = 1}^{{N_{\text{t}}}} {{h_{m,n}^k} s_n^k ( t )}  + {w_m^k}( t ).
       \label{equ_y_mkt_block}
\end{equation}
Here, ${h_{m,n}^k}$ denotes the block-fading channel coefficient between the $n$-th transmit antenna and the $m$-th receive antenna at the $k$-th subcarrier, while ${\mathbf{w}}_m^k = [w_m^k ( t )] \in \mathbb{C}^{T \times 1}$ denotes additive white Gaussian noise (AWGN) with variance $\sigma _w^2$.

The received components at the $k$-th subcarrier can be collected into the matrix ${{\mathbf{Y}}^k} =[{{\mathbf{y}}_{1}^k},\ldots,{{\mathbf{y}}_{N_{\text{r}}}^k}]^{\mathrm{T}} \in \mathbb{C}^{{N_{\text{r}}} \times T}$. Under the block-fading assumption, the MIMO subsystem can be expressed as
\begin{equation}
       \label{equ_yk_mimo}
       {\mathbf{Y}}^k = {\mathbf{H}}^k \left(\sqrt \rho \, {\mathbf{P}} + \alpha \, {\mathbf{D}}^k \left({\mathbf{I}} - {\mathbf{J}} \right) \right) + {\mathbf{W}}^k,
\end{equation}
where ${\mathbf{P}}=[{{\mathbf{p}}_1},\ldots,{{\mathbf{p}}_{N_{\text{t}}}}]^{\mathrm{T}} \in \mathbb{C}^{{N_{\text{t}}} \times T}$, ${\mathbf{D}}^k=[{{\mathbf{d}}_{1}^k},\ldots,{{\mathbf{d}}_{N_{\text{t}}}^k}]^{\mathrm{T}} \in \mathbb{C}^{{N_{\text{t}}} \times T}$, ${\mathbf{H}}^k = [ h_{m,n}^k ] \in \mathbb{C}^{{N_{\text{r}}} \times {N_{\text{t}}}}$ denotes the channel matrix, and ${{\mathbf{W}}^k} \in \mathbb{C}^{{N_{\text{r}}} \times T}$ denotes the collected noise matrix.

\begin{figure*}[t]
       \centering
       \begin{minipage}{0.58\linewidth}
         \centerline{\includegraphics[height=0.94in]{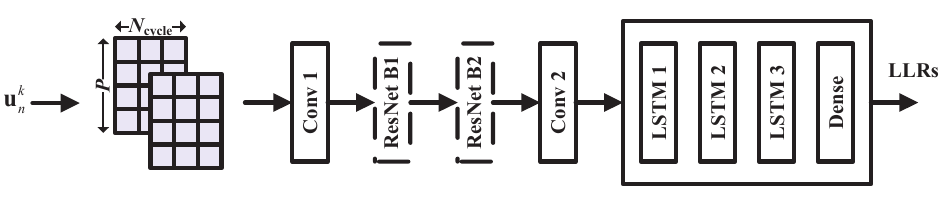}}
         \centerline{\small{(a) Proposed denoiser}}
       \end{minipage}\quad \quad
       \vspace{0.1\baselineskip}
       \begin{minipage}{0.38\linewidth}
         \centerline{\includegraphics[height=0.94in]{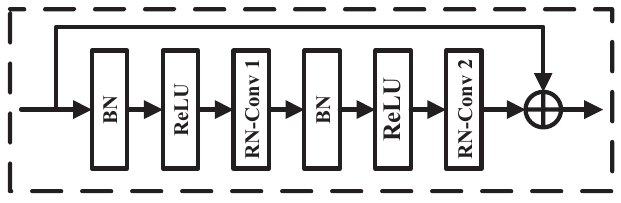}}
         \centerline{\small{(b) ResNet block}}
       \end{minipage}
       \caption{Architecture of the proposed denoising network for the detection module.}
       \label{fig_block_denoiser}
       \vspace*{-1em}
\end{figure*}

\vspace{-1em}
\subsection{Traditional DDST Receiver} 
\label{sec_block_trad}
By utilizing the DDST scheme, the following orthogonal properties hold \cite{DDSTGho}
\begin{equation}
       \left(\mathbf{I}-{\mathbf{J}}\right) {{\mathbf{P}}^{\mathrm{H}}} = \mathbf{0}, ~~ {\mathbf{P}} {{\mathbf{P}}^{\mathrm{H}}} = T\mathbf{I}.
    \label{equ_orthogonal}
\end{equation}
Under the quasi-static channel assumption, the LS-based channel estimate is obtained as
\begin{equation}
       {\mathbf{H}}^{\text{LS},k} = \frac{1}{{T\sqrt {\rho } }} {\mathbf{Y}}^k {{\mathbf{P}}^{\mathrm{H}}} = {\mathbf{H}}^k + \frac{1}{{T\sqrt {\rho} }} {\mathbf{W}}^k {{\mathbf{P}}^{\mathrm{H}}}.
    \label{equ_ls_hk}
\end{equation}
It is observed that the estimation error is independent of the unknown data components, indicating that pilot contamination and data interference are effectively eliminated despite the use of superimposed transmission \cite{DDSTKam}.

A similar interference cancellation procedure is applied in the detection module to obtain the observations associated with the data components:
\begin{align}
       {\mathbf{Z}}^k &= {\mathbf{Y}}^k \left({\mathbf{I}} - {\mathbf{J}} \right) \notag \\
       &= \alpha \, {\mathbf{H}}^k {\mathbf{D}}^k \left({\mathbf{I}} - {\mathbf{J}} \right) + {\mathbf{W}}^k \left({\mathbf{I}} - {\mathbf{J}} \right).
    \label{equ_Zk}
\end{align}
The LMMSE-based detection is then computed as \cite{DDSTGho}
\begin{equation}
       {{\mathbf{U}}^k} = {\left( {{{{\widehat{\mathbf{H}}}}^{\mathrm{H}}}{{\widehat{\mathbf{H}}}} + \left( {1 - \frac{1}{P}} \right)\sigma _w^2{\mathbf{I}}} \right)^{ - 1}}{ {{\widehat{\mathbf{H}}}}^{\mathrm{H}}}{{\mathbf{Z}}^k},
    \label{equ_lmmse_uk}
\end{equation}
where the estimated channel matrix is given by ${\widehat{\mathbf{H}}}=\alpha \, {\mathbf{H}}^{\text{LS},k}$ when the LS-based channel estimation is adopted.

Notably, ${{\mathbf{U}}^k}$ corresponds to the estimate of the distorted data sequence, i.e., ${\mathbf{D}}^k ({\mathbf{I}} - {\mathbf{J}} )$. However, the original data ${\mathbf{D}}^k$ cannot be linearly recovered since $({\mathbf{I}} - {\mathbf{J}} )$ is singular. Therefore, an iterative symbol-by-symbol detection scheme is adopted to obtain the data estimate ${\widehat{\mathbf{D}}}^k$. Specifically, the initial hard decision is computed as
\begin{equation}
       {{\widehat{\mathbf{D}}}^k} = \left\lfloor {{{\mathbf{U}}^k}} \right\rfloor,
    \label{equ_initial_hatdk}
\end{equation}
where $\lfloor {{{\mathbf{U}}^k}} \rfloor$ denotes the matrix obtained by mapping each element of ${{{\mathbf{U}}^k}}$ to its nearest constellation point in Euclidean distance. 
At each subsequent iteration, the detected symbols are updated as \cite{DDSTGho}
\begin{equation}
       {{\widehat{\mathbf{D}}}^k} = \left\lfloor {{\mathbf{U}}^k + {{\widehat{\mathbf{D}}}^k}{\mathbf{J}}} \right\rfloor.
    \label{equ_i_hatdk}
\end{equation}

\vspace{-1em}
\subsection{Enhanced DDST Receiver} 
\label{sec_block_enhance}
Although the orthogonal structure eliminates the reliance on data priors for pilot-data decoupling, conventional DDST receivers still suffer from several inherent limitations. On the one hand, LS-based channel estimation exhibits limited accuracy in low signal-to-noise ratio (SNR) regimes, which subsequently degrades detection performance. On the other hand, the symbol-by-symbol detection strategy is highly sensitive to the initial hard decision and becomes particularly vulnerable to symbol misidentification under high-order modulation. 
Moreover, the hard-decision-based method in \eqref{equ_i_hatdk} cannot provide reliable soft information for transmitted bits, significantly weakening the error correction capability of the channel decoder. As a result, conventional DDST receivers are unsuitable for coded systems. To address these issues, an enhanced DDST receiver is proposed by integrating LMMSE-based channel estimation with a denoising network for signal detection, thereby improving the overall receiver performance.

\subsubsection{Channel Estimation}
According to \eqref{equ_ls_hk}, the LS-based channel estimate can be regarded as a noisy observation of the true channel coefficients. By vectorizing the channel matrix, the observation vector $\widetilde{\mathbf{y}}=\text{vec}({\mathbf{H}}^{\text{LS},k})$ can be expressed as
\begin{equation}
       \widetilde{\mathbf{y}} = \widetilde{\mathbf{h}} + \frac{1}{{T \sqrt \rho  }}\,{\widetilde{\mathbf{w}}},
    \label{equ_observe_yk}
\end{equation}
where ${\widetilde{\mathbf{h}}}=\text{vec}({\mathbf{H}}^{k})$, and ${\widetilde{\mathbf{w}}} = {[ {{\mathbf{p}}_1^{\mathrm{H}}{{\mathbf{w}}_1^k}, \ldots ,{\mathbf{p}}_{{N_{\text{t}}}}^{\mathrm{H}}{{\mathbf{w}}_{{N_{\text{r}}}}^k}} ]^{\mathrm{T}}}$. The Wiener-Hopf equation can then be employed \cite{WHKay}
\begin{subequations}
       \label{equ_WH}
       \begin{align}
       &{\widetilde{\mathbf{h}}}^{\text{LMMSE}} = {\mathbf{C}}_{\mathbf{yh}}^{\mathrm{H}}{\mathbf{C}}_{\mathbf{yy}}^{ - 1}{\widetilde{\mathbf{y}}}, \\
       &{{\mathbf{C}}_{{\mathbf{yh}}}} = \mathbb{E}\left\{ {{\widetilde{\mathbf{y}}}{\widetilde{\mathbf{h}}}^{\mathrm{H}}} \right\} = {{\mathbf{R}}^{{\text{Spat}}}}, \\
       &{{\mathbf{C}}_{{\mathbf{yy}}}} = \mathbb{E}\left\{ {{\widetilde{\mathbf{y}}}{\widetilde{\mathbf{y}}}^{\mathrm{H}}} \right\} = {{\mathbf{R}}^{{\text{Spat}}}} + \frac{\sigma _w^2}{T \rho}{\mathbf{I}},
       \end{align}
\end{subequations}
where ${{\mathbf{R}}^{{\text{Spat}}}} = \mathbb{E}\{\widetilde{\mathbf{h}} {\widetilde{\mathbf{h}}}^{\mathrm{H}} \} \in \mathbb{C}^{{{N_{\text{r}}}{{N_{\text{t}}}}} \times {{N_{\text{r}}}{{N_{\text{t}}}}}}$ denotes the spatial-domain channel correlation matrix. The resulting LMMSE-based channel estimate is obtained as
\begin{equation}
       {\widetilde{\mathbf{h}}}^{\text{LMMSE}}={\left( {{{\mathbf{R}}^{{\text{Spat}}}}} \right)^{\mathrm{H}}}{\left( {{{\mathbf{R}}^{{\text{Spat}}}} + \frac{\sigma _w^2}{T \rho}{\mathbf{I}}} \right)^{ - 1}}{\widetilde{\mathbf{y}}}.
    \label{equ_hlmmsek}
\end{equation} 
The LMMSE-based channel estimate ${\mathbf{H}}^{{\text{LMMSE}},k} \in \mathbb{C}^{{N_{\text{r}}} \times {{N_{\text{t}}}}}$ is then obtained by reshaping ${\widetilde{\mathbf{h}}}^{\text{LMMSE}}$ into matrix form.

\subsubsection{Signal Detection}
Based on the acquired LMMSE-based channel estimate, a similar detection process can be performed according to \eqref{equ_lmmse_uk}, where ${\widehat{\mathbf{H}}}=\alpha \, {\mathbf{H}}^{\text{LMMSE},k}$. By denoting the $n$-th row of $\mathbf{U}^k$ as $\mathbf{u}_n^k = [u_n^k(t)] \in \mathbb{C}^{T \times 1}$, the initial data estimate can be expressed as
\begin{equation}
       {\mathbf{u}}_n^k = {\mathbf{d}}_n^k + {\mathbf{\Delta u}}_n^k.
    \label{equ_uk_input}
\end{equation}
The estimation error ${\mathbf{\Delta u}}_n^k$ consists of both the residual error caused by imperfect CSI in LMMSE filtering and the data-dependent distortion introduced by ${\mathbf{e}}_n^k$. 

To address the intrinsic limitations of conventional detection schemes in terms of demapping reliability and scalability, the initial estimates are fed into a denoising network to generate the log-likelihood ratios (LLRs) of the transmitted bits. The proposed denoiser integrates a batch-normalized (BN) CNN with residual connections based on ResNet blocks \cite{BlockSDHon,SIPAit} and an LSTM network \cite{BlockSDGui,BlockSDEmi} to exploit the temporal correlation of data-dependent distortion, thereby facilitating the recovery of intrinsic data features. The overall architecture of the proposed denoiser is illustrated in \figref{fig_block_denoiser}.

As shown in \figref{fig_block_denoiser}, the initial estimates ${\mathbf{u}}_n^k$ are converted into two feature maps by concatenating their real and imaginary parts before being fed into the denoiser. Furthermore, according to \eqref{equ_J}, the perturbation sequence exhibits periodic similarity:
\begin{equation}
       {e_n^k}\left( t \right) = {{\mathbf{j}}_t^{\mathrm{T}}{{\mathbf{d}}_n^k}} = {e_n^k}\left( {t + p{N_{\text{cycle}}}} \right), ~~ p = 1, \ldots ,P,
    \label{equ_e_period}
\end{equation}
where ${\mathbf{j}}_t$ denotes the $t$-th column of $\mathbf{J}$. Consequently, the input feature maps are reshaped into dimensions of $P \times N_{\text{cycle}}$, enabling the CNN-based architecture to exploit periodic similarity for enhanced denoising. The input features are first processed by $C_1$ convolutional filters of size $F_1 \times F_1 \times 2$ at Conv1, followed by two ResNet blocks. Each ResNet block consists of two batch-normalized convolutional layers with ReLU activation, employing $C_1$ filters of size $F_2 \times F_2 \times C_1$. The extracted features are then processed by Conv2 with $C_1$ filters of size $F_1 \times F_1 \times C_1$, producing an output of size $P \times {N_{\text{cycle}}} \times C_1$.

To further capture temporal correlations in the residual features, the extracted features are reshaped into a $T \times C_1$ representation and processed by a four-layer LSTM-based network. Specifically, the first two layers consist of LSTM units with $C_1$ hidden cells. The third layer employs $M$ LSTM cells to map the extracted data features into the log-probability space. Finally, a fully connected (FC) layer with $Q$ neurons projects the likelihood features to LLRs, resulting in an output of size $T \times Q$.

The quasi-static channel assumption guarantees orthogonality between pilot and distorted data components, thereby facilitating pilot extraction and suppressing data interference during CSI acquisition. In contrast, under time-varying conditions, this orthogonality is disrupted, making pilot-data decoupling fundamentally challenging and causing severe degradation in channel estimation accuracy. The resulting inaccurate CSI further aggravates residual interference in the initial data estimates, making it difficult for the denoising network to reliably map distorted observations to extrinsic bit-level information. Consequently, the current DDST receiver design becomes ineffective.

\begin{figure}[t]
       \centering
       \begin{minipage}{2.5in}
         \centerline{\includegraphics[width=2.5in]{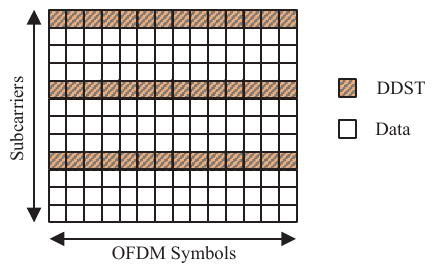}}
       \end{minipage}
       \caption{Illustration of resource allocation within a resource block (RB) under the proposed mix transmission scheme with $r=1/4$ as an example.}
       \label{fig_tv_frame}
       \vspace{-1em}
\end{figure}

\newcounter{TempEqCnt}
\setcounter{TempEqCnt}{\value{equation}}
\setcounter{equation}{21}
\begin{figure*}[!b]
       \centering
       \vspace{-1em}
       \hrulefill
       \begin{equation}
         \label{equ_delta_mnk1t}
         {\delta}_{m,n}^{k}( t ) = p_n^ * ( t )\left[ { \sum\limits_{n' \ne n}^{{N_{\text{t}}}} {{p_{n'}}( t )h_{m,n'}^{{k}}( t )}  + {\frac{\alpha}{\sqrt \rho}} \, \left( {\sum\limits_{n' = 1}^{{N_{\text{t}}}} {d_{n'}^{{k}}( t )h_{m,n'}^{{k}}( t )} - \sum\limits_{n' = 1}^{{N_{\text{t}}}} {{\mathbf{j}}_t^{\mathrm{T}}{\mathbf{d}}_{n'}^{{k}}h_{m,n'}^{{k}}( t )} } \right)} \right]
       \end{equation}
\end{figure*}

\setcounter{equation}{\value{TempEqCnt}}

\section{Proposed Neural Receiver under Time-varying Scenarios}  
\label{sec_tv}
In this section, the loss of orthogonality under time-varying channels is analyzed, based on which a refined transmission scheme is developed to address the superimposed coupling among pilot, data, and data-dependent distortion components. Furthermore, a neural receiver is proposed to recover the underlying orthogonal structure by leveraging the ViT architecture and self-attention mechanism.

\vspace{-1em}
\subsection{Frame Structure and Problem Formulation} 
\label{sec_tv_system}
Under the block-fading assumption, the conventional DDST scheme allocates all time-frequency REs to DDST transmission for estimating the channel frequency responses across subcarriers. In contrast, for time-varying channels, the transmission frame is restructured by partitioning the frequency-domain REs into two segments: one for DDST symbol transmission and the other for pure data transmission. As illustrated in \figref{fig_tv_frame}, the mix transmission scheme uses a parameter $r \in (0,1)$ to denote the proportion of REs allocated to DDST transmission. The transmitted symbol from the $n$-th antenna corresponding to the $(k,t)$-th RE is expressed as
\begin{equation}
       s_n^k ( t ) = \left\{ {\begin{array}{*{20}{l}}
              {\sqrt \rho \, {p_n}( t ) + \alpha \, {\widetilde{d}}_n^k ( t ),}&{k \in \mathcal{K}_1} \\ 
              {d_n^k ( t ),}&{k \in \mathcal{K}_2} 
            \end{array}} \right. ,
       \label{equ_s_nkt_mix}
\end{equation}
where $\mathcal{K}_1$ and $\mathcal{K}_2$ denote the sets of subcarrier indices assigned to DDST transmission and pure data transmission, respectively.

\begin{figure*}[t]
       \centering
       \includegraphics[width=0.8\textwidth]{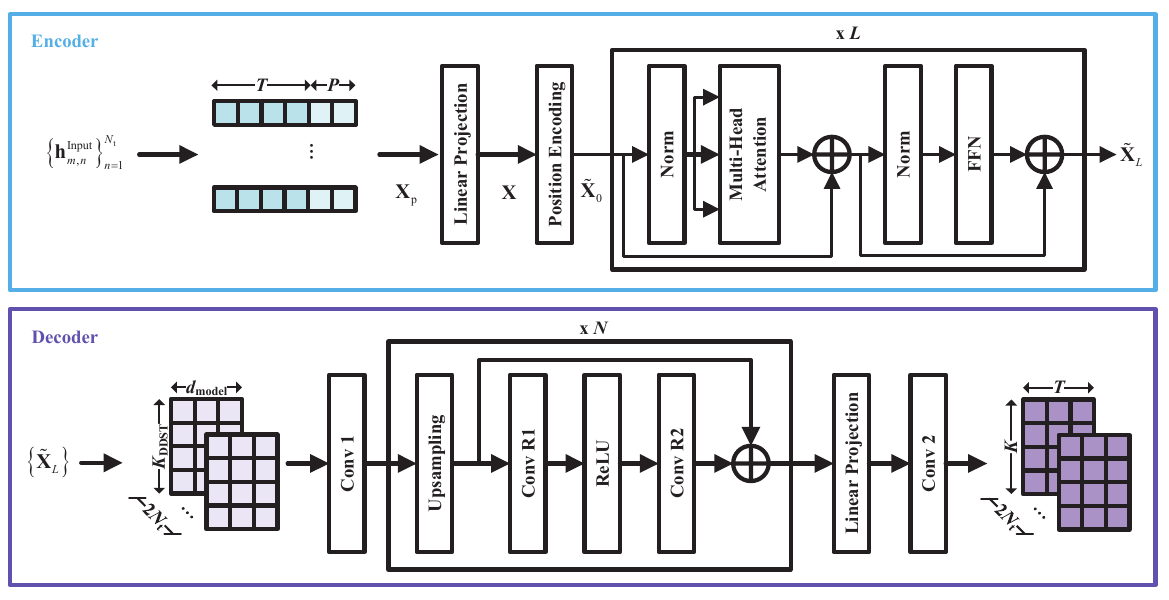}
       \caption{Architecture of the proposed channel estimation network under the mix transmission scheme.}
       \label{fig_tv_vit}
       \vspace*{-1em}
\end{figure*}

Unlike \eqref{equ_y_mkt_block}, the relationship between the received signals and transmitted symbols is given by
\begin{equation}
       y_m^k ( t ) = \sum\limits_{n = 1}^{{N_{\text{t}}}} {{h_{m,n}^k ( t )}  s_n^k ( t )}  + {w_m^k}( t ),
       \label{equ_y_mkt_tv}
\end{equation}
where ${h_{m,n}^k ( t )}$ denotes the time-varying channel coefficient. 
The corresponding MIMO subsystem is expressed as
\begin{equation}
       \label{equ_ykt_mimo_tv}
       {\mathbf{y}}^k ( t ) = {{\mathbf{H}}^k ( t )} {{\mathbf{s}}^k ( t )}  + {\mathbf{w}}^k ( t ),
\end{equation}
where ${\mathbf{y}}^k ( t )=[y_m^k ( t )] \in \mathbb{C}^{{N_{\text{r}}} \times 1}$ collects the received components at the $(k,t)$-th RE from ${N_{\text{r}}}$ antennas. Likewise, ${\mathbf{H}}^k ( t ) = [ h_{m,n}^k ( t ) ] \in \mathbb{C}^{{N_{\text{r}}} \times {N_{\text{t}}}}$ denotes the channel matrix, ${\mathbf{w}}^k ( t )=[w_m^k ( t )] \in \mathbb{C}^{{N_{\text{r}}} \times 1}$ denotes the noise vector, and the transmitted symbol vector ${\mathbf{s}}^k ( t )=[s_n^k ( t )] \in \mathbb{C}^{{N_{\text{t}}} \times 1}$ is defined as
\begin{equation}
       \label{equ_skt_mimo_tv}
       {\mathbf{s}}^k ( t ) = \left\{ {\begin{array}{*{20}{l}}
              {\sqrt \rho \, {\mathbf{p}}( t ) + \alpha \, \left( {{{\mathbf{d}}^k}( t ) - {{\mathbf{e}}^k}( t )} \right),}&{k \in \mathcal{K}_1} \\ 
              {{\mathbf{d}}^k ( t ),}&{k \in \mathcal{K}_2} 
            \end{array}} \right. ,
\end{equation}
where ${\mathbf{p}} ( t )$, ${\mathbf{d}}^k ( t )$, and ${\mathbf{e}}^k ( t )$ collect the pilot, data, and perturbation components, respectively.

As indicated in \eqref{equ_y_mkt_tv}, the time-varying channel response imposes different gains on the transmitted sequence across time instants. Based on the pilot information at the DDST REs, the initial LS-based channel estimate is derived as
\begin{align}
       \label{equ_hls_mnk1t}
       h_{m,n}^{{\text{LS}},{k}}( t ) &= \frac{1}{\sqrt \rho} p_n^ * ( t ) y_m^{k} ( t ) \notag \\
       &= h_{m,n}^{{k}}( t ) + {\delta}_{m,n}^{k}( t ) + \frac{1}{{\sqrt \rho  }} p_n^ * ( t ) w_m^{{k}}( t ),
\end{align}
where $k \in \mathcal{K}_1$, and the residual error ${\delta}_{m,n}^{k}( t )$ is further detailed in \eqref{equ_delta_mnk1t}. Here, $\sum_{\substack{n' \ne n}}$ denotes summation over all transmit antennas except the $n$-th one. It can be observed that the orthogonality between the pilot and perturbation-bearing data components is violated in the received signal. As a result, pilot-data decoupling is no longer supported by LS-based estimation, leading to a complicated interference structure characterized by ${\delta}_{m,n}^{k}(t) \neq 0$. A similar limitation arises in the detection process, where pilot interference cannot be completely removed when forming data observations, unlike the formulation in \eqref{equ_Zk}.

\vspace{-1em}
\subsection{Channel Estimation: ViT}
\label{sec_tv_ce}
To address the complicated interference structure in \eqref{equ_delta_mnk1t}, a channel estimation network based on the self-attention mechanism is proposed. The network exploits implicit channel correlations and latent orthogonal structures embedded in the LS-based input features, thereby mitigating estimation inaccuracies induced by pilot-data coupling interference. The overall architecture consists of a Transformer-based encoder and a CNN-based decoder, as illustrated in \figref{fig_tv_vit}. Note that both the preprocessing of input features and the feature extraction process of the encoder network are implemented independently on each DDST subcarrier, where $k \in \mathcal{K}_1$. Therefore, the superscript $(\cdot)^k$ is omitted in this subsection for notational simplicity.

\subsubsection{Preprocessing}
The input features for the channel estimation network are derived from the initial LS estimates and are specifically designed to capture the distinct characteristics of interference across both temporal and spatial dimensions. According to \eqref{equ_delta_mnk1t}, the residual error consists of three components, namely pilot contamination, data interference, and perturbation-induced distortion, denoted by ${\delta}_{m,n}^{{\text{A}}}( t )$, ${\delta}_{m,n}^{{\text{B}}}( t )$, and ${\delta}_{m,n}^{{\text{C}}}( t )$, respectively
\setcounter{equation}{22}
\begin{subequations}
       \begin{align}
              \label{equ_vit_delta1}
              {\delta}_{m,n}^{{\text{A}}}( t ) &= \sum\limits_{n' \ne n}^{{N_{\text{t}}}} {v_{n,n'}^{\text{a}}( t )h_{m,n'}( t )} , \\
              {\delta}_{m,n}^{{\text{B}}}( t ) &\propto p_n^ * ( t ) {z_m^{{\text{b}}}( t )}, \label{equ_vit_deltab} \\
              {\delta}_{m,n}^{{\text{C}}}( t ) &\propto \sum\limits_{n' = 1}^{{N_{\text{t}}}} {v_{n,n'}^{{\text{c}}}( t )h_{m,n'}( t )} = p_n^ * ( t ) {z_m^{{\text{c}}}( t )}, \label{equ_vit_deltac}
       \end{align}
\end{subequations}
where
\begin{subequations}
       \begin{align}
              \label{equ_vit_delta2}
              v_{n,n'}^{\text{a}}( t ) &= {p_n^ * ( t ){p_{n'}}( t )},~
              {z_m^{{\text{b}}}( t )} = \sum\limits_{n' = 1}^{{N_{\text{t}}}} {d_{n'}( t )h_{m,n'}( t )}, \\
              v_{n,n'}^{{\text{c}}}( t ) &= {p_n^ * ( t ){\mathbf{j}}_t^{\mathrm{T}}{\mathbf{d}}_{n'}},~
              {z_m^{{\text{c}}}( t )} = \sum\limits_{n' = 1}^{{N_{\text{t}}}} {{{\mathbf{j}}_t^{\mathrm{T}}{\mathbf{d}}_{n'}h_{m,n'}( t )}}.
       \end{align}
\end{subequations}
From the temporal perspective, the periodic structures of the pilot and perturbation sequences induce cyclic behavior in $v_{n,n'}^{\text{a}}( t )$ and $v_{n,n'}^{{\text{c}}}( t )$, respectively, as
\begin{subequations}
       \begin{align}
              \label{equ_vit_time}
              v_{n,n'}^{\text{a}}( t ) &= v_{n,n'}^{\text{a}}( t + p N_{\text{cycle}} ), \\
              v_{n,n'}^{{\text{c}}}( t ) &= v_{n,n'}^{{\text{c}}}( t + p N_{\text{cycle}} ),
       \end{align}
\end{subequations}
for $p=1,\ldots,P$. Accordingly, under the assumption that the channel response remains approximately invariant over neighboring $N_{\text{cycle}}$ time resources, summing the initial LS-based estimates over each group of $N_{\text{cycle}}$ adjacent REs in the time domain yields the despreading feature $\widetilde{h}_{m,n}^{{\text{Des}}}( p )$. This feature contains the pilot contamination component $\widetilde{\delta}_{m,n}^{{\text{A}}}( p )$ and the perturbation component $\widetilde{\delta}_{m,n}^{{\text{C}}}( p )$, which satisfy
\begin{equation}
       \widetilde{\delta}_{m,n}^{{\text{A}}}( p ) \to 0, ~~ \widetilde{\delta}_{m,n}^{{\text{C}}}( p ) \propto \sum\limits_{n' = 1}^{{N_{\text{t}}}} {\alpha_{n,n'} \, h_{m,n'}( p N_{\text{cycle}} )},
       \label{equ_vit_time2} 
\end{equation}
where $\alpha_{n,n'}$ is a scaling factor independent of $p$. It is observed that the despreading operation enhances the temporal similarity of residual features across time. The despreading sequence $\widetilde{\mathbf{h}} _{m,n}^{{\text{Des}}}=[\widetilde{h}_{m,n}^{{\text{Des}}}( p )] \in \mathbb{C}^{P \times 1}$ is then fed into the encoder network to facilitate residual denoising.

Additionally, the initial LS-based estimates are also constructed as input features and fed into the encoder network. This design is motivated by the similarity of residual information across the spatial domain. According to \eqref{equ_vit_deltab} and \eqref{equ_vit_deltac}, for $n=1,\ldots,N_{\text{t}}$, both the data interference component ${\delta}_{m,n}^{{\text{B}}}( t )$ and the perturbation component ${\delta}_{m,n}^{{\text{C}}}( t )$ can be regarded as residual terms proportional to $p_n^ * ( t )$. Accordingly, input patches are constructed using LS estimation vectors from multiple transmit antennas, i.e., $\mathbf{h} _{m,n}^{{\text{LS}}}=[h_{m,n}^{{\text{LS}}}( t )] \in \mathbb{C}^{T \times 1}$ for $n=1,\ldots,N_{\text{t}}$. This design enhances the encoder's capability to suppress residual interference and extract the underlying channel correlations.

Based on the above observations, the despreading sequence and initial estimation vector corresponding to the $(m,n)$-th antenna pair are concatenated to form the input vector
\begin{equation}
       \label{equ_vit_input}
       {\mathbf{h} _{m,n}^{{\text{Input}}}} = \left[\left({\mathbf{h} _{m,n}^{{\text{LS}}}}\right)^{\mathrm{T}}, \left({\widetilde{\mathbf{h}} _{m,n}^{{\text{Des}}}}\right)^{\mathrm{T}} \right]^{\mathrm{T}}.
\end{equation}

\subsubsection{Encoder}
The $N_{\text{t}}$ input vectors, $\{\mathbf{h} _{m,n}^{{\text{Input}}}\}$ for $n=1,\ldots,N_{\text{t}}$, are further processed by separating their real and imaginary parts and reshaped into the flattened 2D patches $\mathbf{X}_{\text{p}} \in \mathbb{R}^{2N_{\text{t}} \times (T+P)}$. Here, $(T+P)$ denotes the length of each flattened patch, while $2N_{\text{t}}$ represents the total number of patches. 
As shown in \figref{fig_tv_vit}, the patches are first mapped into a $d_{\text{model}}$-dimensional space via a trainable linear projection, yielding patch embeddings $\mathbf{X} \in \mathbb{R}^{2N_{\text{t}} \times d_{\text{model}}}$. The patch embeddings are then passed through a positional encoding layer to preserve the relative positional information of channel observations along the time dimension
\begin{equation}
       \label{equ_vit_patch}
       \mathbf{X} = \text{Dense}\left( \mathbf{X}_{\text{p}} \right), ~~ {\widetilde{\mathbf{X}}}_0 = \mathbf{X} + \mathbf{E}_{\text{pos}}.
\end{equation}
Here, $\text{Dense}(\cdot)$ denotes the FC layer used for linear projection, and $\mathbf{E}_{\text{pos}} \in \mathbb{R}^{2N_{\text{t}} \times d_{\text{model}}}$ is the learnable position embeddings. 
The resulting embedding vectors serve as input to $L$ encoding modules, each comprising a multi-head
self-attention (MSA) block and a feed-forward network (FFN). The $l$-th encoding module ($l=1,\ldots,L$) can be expressed by \cite{TvCEDos, TvCELiu}
\begin{subequations}
       \label{equ_vit_msa_ffn}
       \begin{align}
       {\widetilde{\mathbf{X}}}_l^{{\text{MSA}}} &= {\text{MSA}}\left( {{\text{LN}}\left( {\widetilde{\mathbf{X}}}_{l-1} \right)} \right) + {\widetilde{\mathbf{X}}}_{l-1}, \\
       {\widetilde{\mathbf{X}}}_l &= {\text{FFN}}\left( {{\text{LN}}\left( {\widetilde{\mathbf{X}}}_l^{{\text{MSA}}} \right)} \right) + {\widetilde{\mathbf{X}}}_l^{{\text{MSA}}}, 
       \end{align}
\end{subequations}
where layer normalization (LN) is applied before each block and residual connections are employed after each block.

In the MSA block, the query, key, and value matrices are obtained from the input sequences via separate FC layers, denoted as $\mathbf{Q}$, $\mathbf{K}$, and $\mathbf{V}$, respectively. The query matrix $\mathbf{Q}$ can be further separated into $H$ heads, i.e., $\mathbf{Q}=[\mathbf{Q}_1,\ldots,\mathbf{Q}_H]$, with the $h$-th sub-matrix $\mathbf{Q}_h$ having a size of ${2N_{\text{t}} \times \frac{d_{\text{model}}}{H}}$. A similar separate strategy is applied to $\mathbf{K}$ and $\mathbf{V}$, with the $h$-th sub-matrix represented by $\mathbf{K}_h$ and $\mathbf{V}_h$, respectively. At the $h$-th head, the scaled dot-product attention is employed to compute attention scores over input features and adjust the corresponding feature weights \cite{TvCEVas, SIPZou}
\begin{equation}
       \label{equ_vit_msa2}
       \text{SA}_h = \text{softmax} \left( \frac{\mathbf{Q}_h \mathbf{K}_h^{\mathrm{T}}}{\sqrt{d_{\text{model}}/H}} \right) \mathbf{V}_h.
\end{equation}
The outputs from the self-attention heads are concatenated and subsequently mapped through an FC layer to generate the output of $\text{MSA}(\cdot)$. 
As for the FFN block, a multi-layer perceptron (MLP) is adopted, consisting of two FC layers with $\{2d_{\text{model}}, d_{\text{model}}\}$ neurons, where the GELU activation function is applied between them.

Overall, the proposed encoder network exploits the self-attention mechanism to capture latent orthogonal structures within high-dimensional observation sequences, thereby enabling pilot information extraction and suppressing interference induced by data superposition. The capability to model long-range dependencies further enhances the characterization of time-varying channel statistics, which facilitates feature extraction in dynamic scenarios. In addition, the preprocessing stage is designed based on the temporal and spatial characteristics of channel observations, thereby improving both residual suppression and channel feature extraction. The output of the encoder, ${\widetilde{\mathbf{X}}}_L$, is subsequently passed to the decoder network for further processing.

\subsubsection{Decoder}
Since the Transformer-based design enables the encoder to focus on salient features, its output captures the correlated structure of channels based on observations at the DDST REs. However, the self-attention mechanism primarily emphasizes global interactions among token embeddings, while local structural information is largely overlooked \cite{TvCEYuan, TvCELi}. Moreover, interpolation in the frequency domain is required to obtain channel estimates corresponding to pure data REs. Therefore, a decoder network is introduced, adopting a CNN-based architecture to enhance local feature fusion and perform channel extrapolation.

As shown in \figref{fig_tv_vit}, the encoder's outputs across DDST REs, $\{{\widetilde{\mathbf{X}}}_L\}$ for $k \in \mathcal{K}_1$, are collected and reshaped into $2N_{\text{t}}$ feature maps with dimension $K_{\text{DDST}} \times d_{\text{model}}$, where $K_{\text{DDST}}=rK$ denotes the number of DDST subcarriers. These feature maps are subsequently processed by a residual convolutional network \cite{TvCELuan}, which consists of an input convolutional layer, $N$ interpolation modules, a linear projection layer, and an output convolutional layer.

Specifically, the input convolutional layer employs $C_2$ filters with a kernel size of $F_3 \times F_3 \times 2N_{\text{t}}$. Each interpolation module comprises an upsampling layer and a residual block. The nearest-neighbor interpolation \cite{TvCEZhao} is adopted for frequency-domain interpolation with an upsampling factor of 2. The residual block refines the interpolation by capturing local correlations in both time and frequency domains and consists of two convolutional layers with a ReLU activation in between, using $C_2$ filters of size $F_3 \times F_3 \times C_2$. After interpolation, the features of size $K \times d_{\text{model}} \times C_2$ are projected through an FC layer to obtain channel representations across time instants, yielding a dimension of $K \times T \times C_2$. Finally, the output convolutional layer with $2N_{\text{t}}$ filters is applied to generate the real and imaginary components of the refined channel estimates.

\begin{figure*}[t]
       \centering
       \begin{minipage}{0.58\linewidth}
         \centerline{\includegraphics[height=0.81in]{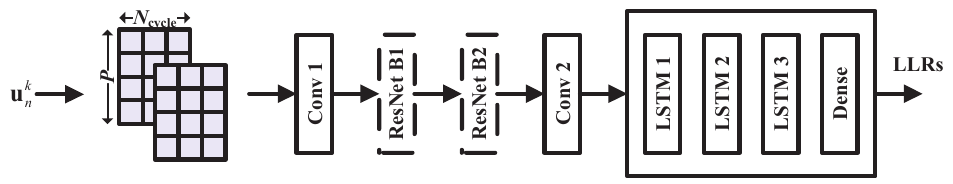}}
         \centerline{\small{(a) Subnet 1 for the DDST REs}}
       \end{minipage}\quad \quad
       \vspace{0.1\baselineskip}
       \begin{minipage}{0.38\linewidth}
         \centerline{\includegraphics[height=0.81in]{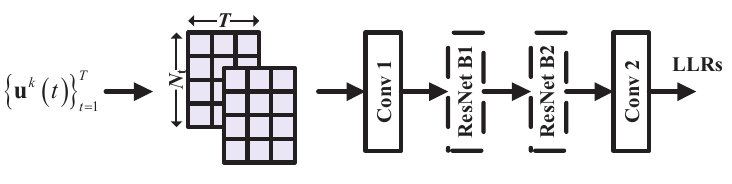}}
         \centerline{\small{(b) Subnet 2 for the Data REs}}
       \end{minipage}
       \caption{Architecture of the proposed detection network under the mix transmission scheme.}
       \label{fig_tv_subnet}
       \vspace*{-1em}
\end{figure*}

\newcounter{TempEqCnt2}
\setcounter{TempEqCnt2}{\value{equation}}
\setcounter{equation}{33}
\begin{figure*}[!b]
       \centering
       \vspace{-1em}
       \hrulefill
       \begin{equation}
              \label{equ_loss_sd}
              {\mathcal{L}_{{\text{SD}}}} = \frac{1}{{{N_{\text{t}}} KTQ}} {\sum\limits_{n = 1}^{{N_{\text{t}}}} {\sum\limits_{k = 1}^{K} {\sum\limits_{t = 1}^{T} {\sum\limits_{q = 1}^{Q}
              {\bigg( b_{n,q}^k(t)\log \Big( {\sigma \big( {L_{n,q}^k( t )} \big)} \Big) + \left( {1 - b_{n,q}^k(t)} \right)\log \Big( {1 - \sigma \big( {L_{n,q}^k( t )} \big)} \Big) \bigg)} 
              } } } } 
       \end{equation}
\end{figure*}
\setcounter{equation}{\value{TempEqCnt2}}

\vspace{-1em}
\subsection{Signal Detection: Hybrid CNN-LSTM}
\label{sec_tv_sd}
High-resolution CSI is obtained at the output of the proposed channel estimation network, where ${{\widehat{h}}_{m,n}^k(t)}$ denotes the estimated channel response corresponding to the $m$-th receive antenna and the $n$-th transmit antenna at the $(k,t)$-th RE. The data observations are then computed, wherein pilot interference is explicitly suppressed from the received signal at the DDST REs
\begin{equation}
       \label{equ_lmmse_zkt_mix}
       {{\mathbf{z}}^k(t)} = \left\{ {\begin{array}{*{20}{l}}
              {{\mathbf{y}}^k (t) - {{\widehat{\mathbf{H}}}^k (t)}( {\sqrt \rho  \, {{\mathbf{p}}(t)}} ),}&{k \in \mathcal{K}_1} \\ 
              {{\mathbf{y}}^k (t),}&{k \in \mathcal{K}_2} 
            \end{array}} \right. ,
\end{equation}
Here, ${\widehat{\mathbf{H}}}^k(t)=[{{\widehat{h}}_{m,n}^k(t)}] \in \mathbb{C}^{N_{\text{r}} \times N_{\text{t}}}$ refers to the estimated channel matrix collected at each RE. Accordingly, the initial LMMSE-based detection can be performed
\begin{subequations}
       \label{equ_lmmse_ukt_mix}
       \begin{align}
       {{\mathbf{u}}^k}(t) &= {\left( {{{{\widehat{\mathbf{H}}}}^{\mathrm{H}}}{{\widehat{\mathbf{H}}}} + \sigma _w^2{\mathbf{I}}} \right)^{ - 1}}{ {{\widehat{\mathbf{H}}}}^{\mathrm{H}}}{{\mathbf{z}}^k(t)}, \\
       {\widehat{\mathbf{H}}} &= \left\{ {\begin{array}{*{20}{l}}
              {\alpha \, {\widehat{\mathbf{H}}}^k (t),}&{k \in \mathcal{K}_1} \\ 
              {{\widehat{\mathbf{H}}}^k (t),}&{k \in \mathcal{K}_2} 
            \end{array}} \right. ,
       \end{align}
\end{subequations}
with $u_n^k(t)$ representing the $n$-th element of ${{\mathbf{u}}^k}(t)$. 
At the DDST REs, the residual structure in the initial estimates arises from imperfect pilot interference cancellation in \eqref{equ_lmmse_zkt_mix} and residual errors of the LMMSE filtering, both of which are caused by imperfect CSI. In addition, the data-dependent distortion introduced by ${\mathbf{e}}^{k}(t)$ further degrades the reliability of data detection and demapping. As for the REs corresponding to pure data transmission, the residual errors are solely attributed to imperfect CSI in the filtering process.

Consequently, two dedicated subnetworks are designed to perform denoising and symbol demapping for different RE segments, as illustrated in \figref{fig_tv_subnet}. For each DDST subcarrier $k \in \mathcal{K}_1$, the initial estimates ${{\mathbf{u}}_n^{k}}=[u_n^{k}(t)] \in \mathbb{C}^{T \times 1}$ are processed by Subnet 1 to recover the intrinsic data features. Subnet 1 employs the ResNet blocks facilitated by the LSTM-based network, which is identical to that used in the denoiser in \secref{sec_block_enhance}.

As for the data REs $k \in \mathcal{K}_2$, since the interference structure in the initial estimates does not contain perturbation components, Subnet 2 focuses on the ResNet-based architecture to suppress filtering noise, while omitting the LSTM-based temporal modeling of data-dependent distortion used in Subnet 1. According to \figref{fig_tv_subnet}(b), the length-$N_{\text{t}}$ initial estimation vectors at different time instants, i.e., $\{{{\mathbf{u}}^{k}}(t)\}$ for $t=1,\ldots,T$, are collected and reshaped into two feature maps with a dimension of $N_{\text{t}} \times T$, corresponding to the real and imaginary parts, respectively. Subnet 2 consists of an input convolutional layer followed by two sequential ResNet blocks, with convolutional configurations $C_1$, $F_1$, and $F_2$ identical to those used in the denoiser. The resulting feature maps of size $N_{\text{t}} \times T \times C_1$ are finally processed by $Q$ filters at the output convolutional layer to produce bit-level representations.

The extrinsic LLRs for the $q$-th bit of the transmitted data symbol corresponding to the $n$-th transmit antenna at the $(k,t)$-th RE, $L_{n,q}^{k}(t)$, can be collected at the output of the proposed detection network. During inference, these LLRs are subsequently delivered to the channel decoder to enable error correction in coded systems.

\vspace{-1em}
\subsection{Training Strategy}
\label{sec_tv_train}
In this subsection, the training procedure of our proposed neural receiver is described in detail, which comprises three stages. In the first stage, the channel estimation network is trained in a supervised manner for 300 epochs. The objective function, denoted by ${\mathcal{L}_{{\text{CE}}}}$, is defined as the mean squared error (MSE) between the output refined channel estimates and the corresponding ground-truth channel coefficients
\begin{equation}
       \label{equ_loss_ce}
       {\mathcal{L}_{{\text{CE}}}} = \frac{1}{{{N_{\text{r}}}{N_{\text{t}}} KT}} {\sum\limits_{m = 1}^{{N_{\text{r}}}} {\sum\limits_{n = 1}^{{N_{\text{t}}}} {\sum\limits_{k = 1}^{K} {\sum\limits_{t = 1}^{T}
       {{{\left| {{\widehat{h}}_{m,n}^k(t)} - {h_{m,n}^k(t)} \right|}^2}} 
       } } } } .
\end{equation}
The network parameters are optimized using the Adam optimizer with a learning rate of $10^{-3}$, where a weight decay of $1\times10^{-4}$ is applied.

In the second stage, the pre-trained channel estimation network is incorporated with its parameters fixed. Based on the estimated CSI, the input features for the detection network are constructed. The detection network is subsequently trained using the Adam optimizer with a learning rate of $10^{-3}$ over 500 epochs. The loss function, denoted by ${\mathcal{L}_{{\text{SD}}}}$, is formulated as the binary cross-entropy between the predicted LLRs and the corresponding transmitted bit sequences, as represented by \eqref{equ_loss_sd}.\footnote{Note that the same training strategy in \eqref{equ_loss_sd} is also adopted for the proposed denoiser in \secref{sec_block_enhance}.} 
Here, $b_{n,q}^k(t)$ refers to the $q$-th bit transmitted in the $(k,t)$-th RE associated with the $n$-th transmit antenna, and $\sigma (x)=1/(1+\exp(-x))$ is the sigmoid function.

In the final stage, both the channel estimation and detection models are jointly fine-tuned in an end-to-end manner. The overall training objective is defined as a weighted combination of the two objectives
\setcounter{equation}{34}
\begin{equation}
       \label{equ_loss_tune}
       {\mathcal{L}} = \lambda {\mathcal{L}_{{\text{CE}}}} + \left( 1 - \lambda \right) {\mathcal{L}_{{\text{SD}}}},
\end{equation}
where $\lambda \in (0,1)$ is a weighting coefficient that balances channel estimation accuracy and detection performance, and is chosen as $\lambda=0.2$. The joint optimization is conducted for 300 epochs using the Adam optimizer with a learning rate of $10^{-4}$, thus enhancing the overall inference performance of the proposed design.

\begin{figure*}[t]
       \centering
       \begin{minipage}{0.46\linewidth}
         \centerline{\includegraphics[height=1.4in]{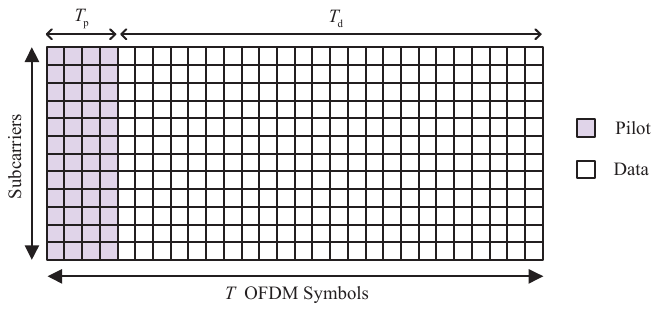}}
         \centerline{\small{(a) $T_\text{p}$ out of $T$ symbols used for pilot transmission}}
       \end{minipage}\quad
       \vspace{0.1\baselineskip}
       \begin{minipage}{0.5\linewidth}
         \centerline{\includegraphics[height=1.4in]{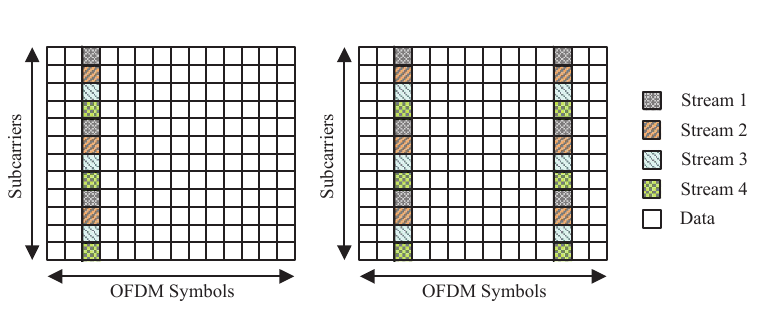}}
         \centerline{\small{(b) 1P and 2P pilot patterns from 5G NR}}
       \end{minipage}
       \caption{Illustration of the pilot placement strategy used in the OP transmission scheme.}
       \label{fig_op_pattern}
       \vspace*{-1em}
     \end{figure*}

\vspace{-1em}
\section{Simulation Results}
\label{sec_simulation}
In this section, numerical results are presented. First, the simulation setup for the MIMO-OFDM system is introduced. We then consider both the quasi-static and time-varying scenarios, and evaluate the performance of the proposed design in comparison with baseline receiver schemes. A complexity analysis is also provided to highlight the computational efficiency improvement enabled by the incorporation of the data-dependent structure. Finally, the generalization capability of the proposed neural receiver under the mix transmission scheme is examined across various channel conditions.

\vspace{-1em}
\subsection{Parameter Settings}
\label{sec_simu_parameter}
An uplink MIMO-OFDM system is simulated with $N_{\text{r}}=16$ receive antennas at the base station (BS) and $N_{\text{t}}=4$ transmit antennas at the user equipment (UE). The transmission frame comprises $T=28$ consecutive OFDM symbols and $K=72$ subcarriers, corresponding to two time slots per frame. 
In both the DDST and mix transmission schemes, all time-frequency REs are available for data transmission. Considering a coded case, the low-density parity-check (LDPC) encoder maps $k=2016$ information bits into coded bits at a rate of $1/2$ for each transmit antenna. The resulting codewords are subsequently modulated using 16-QAM, and the corresponding complex symbols are mapped onto $N_{\text{RB}}=6$ RBs to form the transmitted subframe at each time slot.

\begin{table}[t]
       \caption{Parameters for the Simulation}
       \label{tab_mimo_ofdm}
       \centering
       \begin{tabular}{ll}
         \toprule
         Parameter & Value \\
         \midrule
         Antennas & $16 \times 4$ \\
         Time-frequency REs & $K=72,~T=28$ \\
         Power Delay Profile & CDL-C \\
         Carrier Frequency & 2 GHz \\
         Subcarrier Spacing & 30 kHz \\
         Delay Spread & $\text{ds} \in \{93,363\}$ ns \\
         Testing UE Speed & $v \in \{0,36,108\}$ km/h \\
         Modulation & 16-QAM \\
         Channel Coding Scheme & LDPC $(2016,~1/2)$ \\
         Power Allocation & $\rho \in \{1/7,0.3\}$ \\
         Resource Allocation & $r \in \{1/2,1/4,1/8\}$ \\
         \bottomrule
       \end{tabular}
       \vspace{-1em}
     \end{table}

Channel propagation is modeled using the clustered delay line (CDL) channel model with the CDL-C power delay profile \cite{CDLCite}, as elaborated in \tabref{tab_mimo_ofdm}. The urban macrocell (UMa) scenario is considered, with a delay spread set of $\text{ds} \in \{93,363\}$ ns employed to simulate different frequency-selective fading properties of channels. Additionally, the Doppler effect is characterized by UE velocities $v$, with $v=0$ km/h set for the quasi-static case, and $v \in \{36,108\}$ km/h are employed to simulate time-varying fading properties.

The hyperparameter configurations of the neural models are specified as follows. In the channel estimation module, the ViT-based encoder employs $d_{\text{model}}=128$, $L=2$, and $H=4$. The CNN-based decoder is configured with $C_2 = 16$ and $F_3 = 5$, whereas $N$ varies with the resource allocation ratio $r$. Furthermore, the hybrid CNN-LSTM architecture adopts $C_1=32$, $F_1=4$, and $F_2=3$, and is employed in both the denoiser of the enhanced DDST receiver and the detection subnetworks of the proposed neural receiver. 
The training dataset contains $2.56\times10^5$ MIMO-OFDM channel samples, with 80\% used for training and 20\%
used for validation. A batch size of 16 is adopted. The channel parameters of the test datasets are selected according to the corresponding scenarios, and each test point contains 4800 channel samples. All simulations are implemented using the NVIDIA Sionna package \cite{SionnaHoy}. 
Note that the models evaluated in \secref{sec_simu_block} and \secref{sec_simu_tv} are trained on channel datasets generated under conditions matched to the target test scenarios. In contrast, \secref{sec_simu_mm} examines the direct deployment of pretrained models under mismatched testing conditions, which are explicitly labeled as ``Mismatch.''

\begin{figure}[t]
       \centering
       \begin{minipage}{3in}
         \centerline{\includegraphics[width=3in]{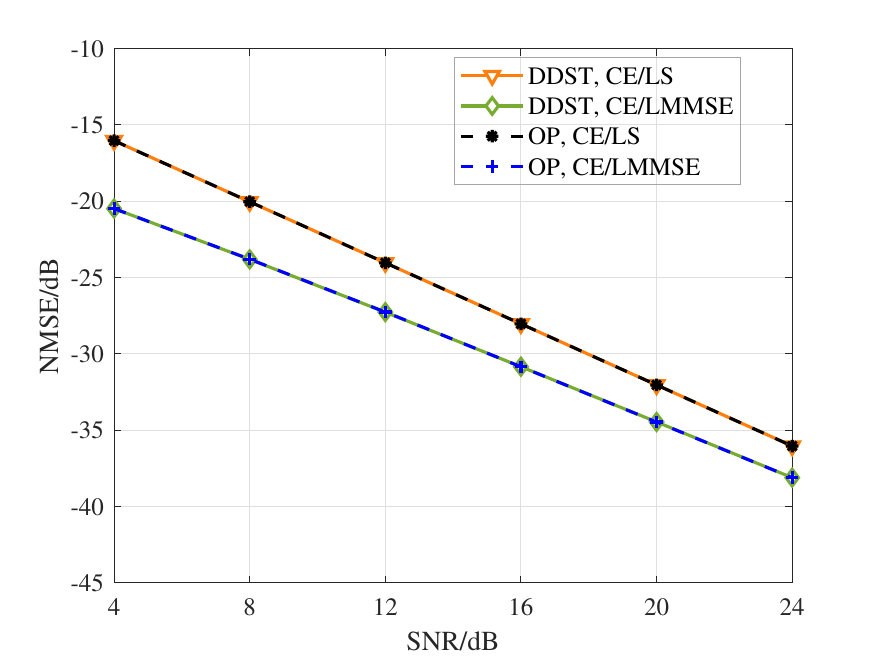}}
         \centerline{\small{(a) NMSE of channel estimation}}
       \end{minipage}
       \hfill
       \begin{minipage}{3in}
         \centerline{\includegraphics[width=3in]{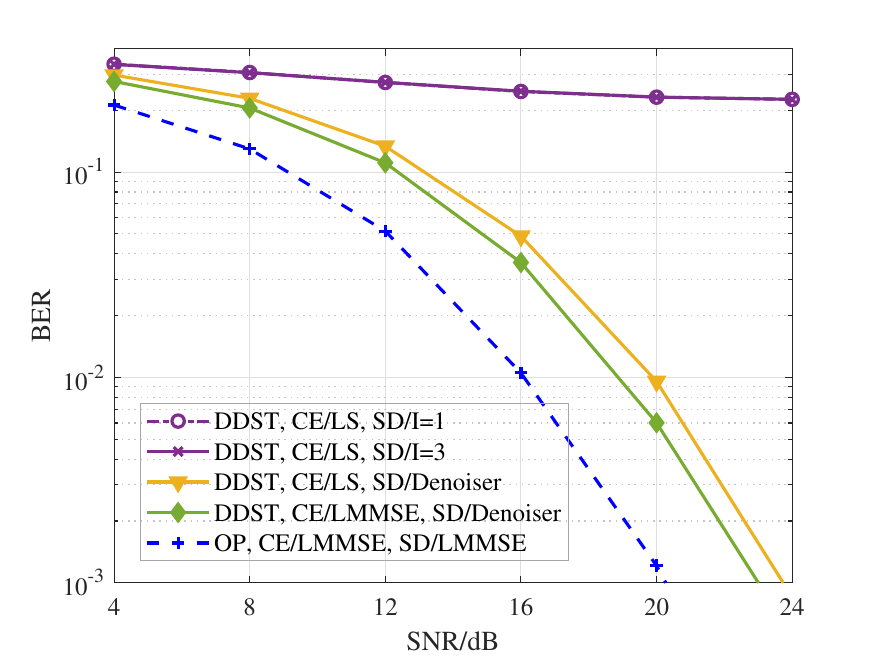}}
         \centerline{\small{(b) BER of symbol demapping}}
       \end{minipage}
       \caption{Performance comparisons between the developed enhanced DDST receiver and the baseline receivers when $T_\text{p}=4,~\rho=1/7$ are set under the block-fading case.}
       \label{fig_block_comp}
       \vspace{-1em}
     \end{figure}

\vspace{-1em}
\subsection{Performance Analysis under Block-fading Scenarios}
\label{sec_simu_block}
In this subsection, the quasi-static assumption of channels is considered, and the performance of the developed enhanced DDST receiver is evaluated. To highlight the effectiveness of the LMMSE-based derivation and the proposed denoiser, the channel estimation accuracy and detection performance are evaluated separately. The benchmark schemes include the OP-based receiver and the conventional DDST receiver introduced in \secref{sec_block_trad}. For the OP baseline, pilot symbols are time-division multiplexed (TDM) with the data. As shown in \figref{fig_op_pattern}(a), $T_\text{p}=4$ out of $T$ OFDM symbols are allocated to pilot transmission from different transmit antennas, while the remaining $T_\text{d}=T-T_\text{p}$ symbols are used for data transmission. For the conventional DDST receiver, the number of iterations with respect to the symbol-by-symbol detection is set to $I\in \{1,3\}$. The power allocation factor for the DDST scheme is set to $\rho = 1/7$.
 
The normalized MSE (NMSE) performance comparisons between the OP receivers and the DDST receivers are exhibited in \figref{fig_block_comp}(a). Both the LS-based and LMMSE-based channel estimates are evaluated, denoted as ``CE/LS'' and ``CE/LMMSE,'' respectively. Theoretically, when $\rho=T_\text{p} / T$ is satisfied, the MSE performance of the OP-based channel estimation and the DDST-based counterpart is equivalent \cite{DDSTGho}. \figref{fig_block_comp}(a) validates this theoretical equivalence, demonstrating effective pilot-data separation in DDST receivers despite the superimposed transmission structure. Moreover, the derived LMMSE-based channel estimation in the enhanced DDST receiver achieves a significant performance gain over the LS-based estimator in conventional DDST designs, thereby improving CSI acquisition accuracy.

Subsequently, the bit error rate (BER) performances between the proposed enhanced DDST receiver and the baseline schemes are presented in \figref{fig_block_comp}(b). Note that the proposed denoiser, labeled ``SD/Denoiser,'' directly outputs bit-level features during inference, making it well suited for coded systems. In contrast, the data estimates obtained by the baseline schemes require additional processing. Specifically, the OP-based receiver employs a parallel interference cancellation (PIC)-based LLR computation method \cite{MMSEICStu}, denoted as ``SD/LMMSE.'' For the conventional DDST receiver, soft demapping is directly performed on $\widehat{\mathbf{D}}^k$, corresponding to the cases labeled ``SD/\emph{I}=1'' and ``SD/\emph{I}=3.''

According to the figure, the symbol-by-symbol detection scheme fails to provide effective iterative refinement under the 16-QAM modulation. This is because the initial hard decision in \eqref{equ_initial_hatdk} suffers from severe symbol misidentification, and the resulting erroneous feedback cannot effectively suppress the perturbation components in subsequent iterations. Moreover, the hard decision-based mechanism cannot provide reliable extrinsic information to support the error correction capability of the channel decoder. In contrast, the proposed denoiser overcomes these shortcomings and delivers substantial performance improvements compared with the conventional DDST receiver.

However, the current denoising network still exhibits notable limitations. Specifically, the hybrid CNN-LSTM architecture cannot fully suppress the intrinsic data-dependent distortion. As evidenced in \figref{fig_block_comp}, although the enhanced DDST receiver achieves channel estimation accuracy comparable to that of the OP baseline, a significant performance gap persists in demapping performance. This residual symbol misidentification can further propagate into the bit-mapping stage, thereby constraining the channel decoder from fully exploiting its error correction potential. 
Nevertheless, the DDST receiver design retains the spectral efficiency advantage over the OP-based scheme. Moreover, simulation results in later subsections demonstrate that our proposed mix transmission scheme effectively alleviates this insufficient distortion suppression, leading to a pronounced improvement in demapping performance.

\vspace{-1em}
\subsection{Performance Analysis under Time-varying Scenarios}
\label{sec_simu_tv}

In this subsection, the performance of the proposed neural receiver introduced in \secref{sec_tv} is evaluated under time-varying channel conditions. Specifically, a UE velocity of $v=108$ km/h and a delay spread of $\text{ds}=363$ ns are considered to characterize the temporal and frequency selectivity of the channel. Two representative baselines are included for comparison, namely the conventional OP receiver and a state-of-the-art SIP receiver. The corresponding simulation configurations are summarized as follows:
\begin{itemize}
  \item \textbf{OP, 2P}: The OP baseline adopts the 5G new radio (NR) pilot pattern illustrated in \figref{fig_op_pattern}(b), where the ``2P'' configuration is selected to match the medium-mobility scenario. The LMMSE-based channel estimation and detection are employed \cite{LMMSECite}, along with the PIC-based soft demapping \cite{MMSEICStu} identical to that described in \secref{sec_simu_block}.
  \item \textbf{SIP, \emph{I}=1 or \emph{I}=2 or \emph{I}=4}: The iterative SIP receiver in \cite{SIPLi} is considered, with the number of JCDD iterations set to $I \in \{1,2,4\}$. Pilot symbols are superimposed over all REs with a power allocation ratio of $\rho = 0.3$, and the LMMSE-based channel estimation is applied.
  \item \textbf{Mix(\emph{r}=1/2 or \emph{r}=1/4 or \emph{r}=1/8)}: The developed neural receiver, featuring the self-attention-based channel estimation network and the hybrid CNN-LSTM-based detection subnetworks, is evaluated under the proposed mix transmission scheme. The resource allocation ratio is set to $r \in \{\frac{1}{2},\frac{1}{4},\frac{1}{8}\}$, and the number of interpolation modules within the decoder of channel estimation network is correspondingly configured as $N \in \{1,2,3\}$, respectively. The power allocation ratio for the DDST REs is fixed to $\rho = 0.3$ according to our empirical observations, and is consistently used throughout the subsequent evaluations.
\end{itemize}

\begin{figure}[t]
       \centering
       \begin{minipage}{3in}
         \centerline{\includegraphics[width=3in]{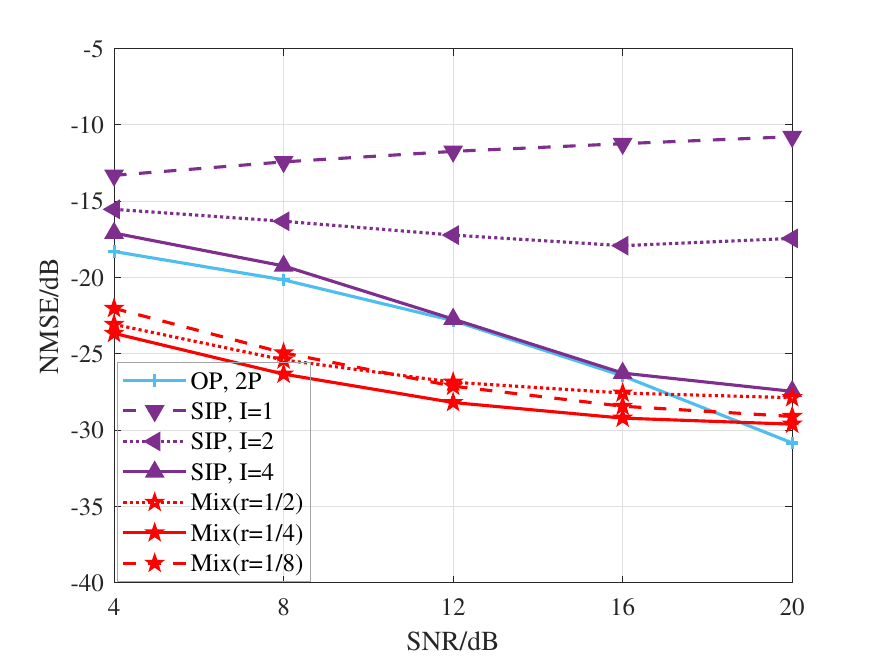}}
         \centerline{\small{(a) NMSE of channel estimation}}
       \end{minipage}
       \hfill
       \begin{minipage}{3in}
         \centerline{\includegraphics[width=3in]{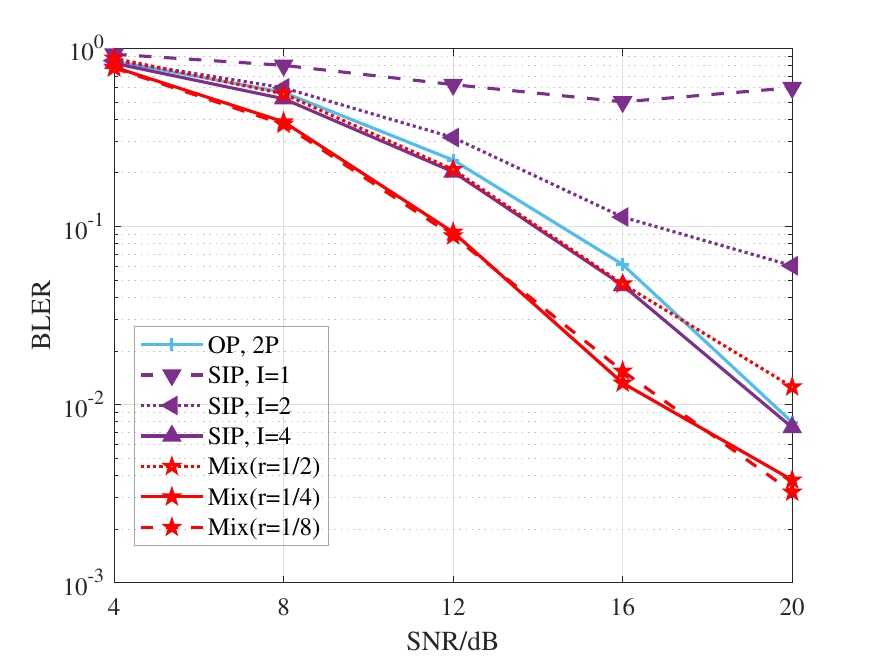}}
         \centerline{\small{(b) BLER of symbol demapping}}
       \end{minipage}
       \caption{Performance comparisons between our proposed neural receiver and the baseline receivers under $v=108$ km/h and $\text{ds}=363$ ns.}
       \label{fig_tv_comp}
     \end{figure}

\figref{fig_tv_comp} presents the channel estimation performance and demapping results in terms of block error rate (BLER). For the SIP receiver, the complex superimposed structure in multi-stream transmissions necessitates multiple JCDD iterations (up to $I=4$) to iteratively refine the a priori information, enabling performance comparable to the OP baseline. This, however, incurs a notable increase in computational latency. In contrast, the proposed receiver integrates the perturbation-based design with the Transformer-based encoder to directly extract pilot information, thereby removing the need for iterative feedback. As shown in \figref{fig_tv_comp}(a), the proposed receiver achieves superior NMSE performance under the mix transmission scheme while significantly reducing runtime, as demonstrated in \tabref{tab_complexity}. Furthermore, even when the resource allocation ratio decreases from $r=\frac{1}{2}$ to $r=\frac{1}{8}$, corresponding to reduced pilot density, the estimation accuracy remains largely unchanged. This observation suggests that the considered pilot densities are sufficient to capture the frequency-selective fading characteristics, and that reliable channel interpolation is effectively achieved by the proposed CNN-based decoder.

In contrast to the comparable channel estimation accuracy observed across different values of $r$, a notable BLER improvement is achieved when employing the mix scheme with $r \in \{\frac{1}{4},\frac{1}{8}\}$ compared to the case of $r=\frac{1}{2}$, as illustrated in \figref{fig_tv_comp}(b). This phenomenon can be explained by the presence of residual data-dependent distortion at the DDST REs, as discussed in \secref{sec_simu_block}, which introduces bit-level misidentification. As $r$ increases, a larger fraction of REs is subject to such distortion, resulting in degraded LLR reliability and exacerbated error propagation during channel decoding. Hence, when CSI acquisition quality is comparable, a smaller $r$, corresponding to a higher proportion of perturbation-free data transmissions, yields more reliable LLRs and consequently improves demapping performance.

\begin{figure}[t]
       \centering
       \begin{minipage}{3in}
         \centerline{\includegraphics[width=3in]{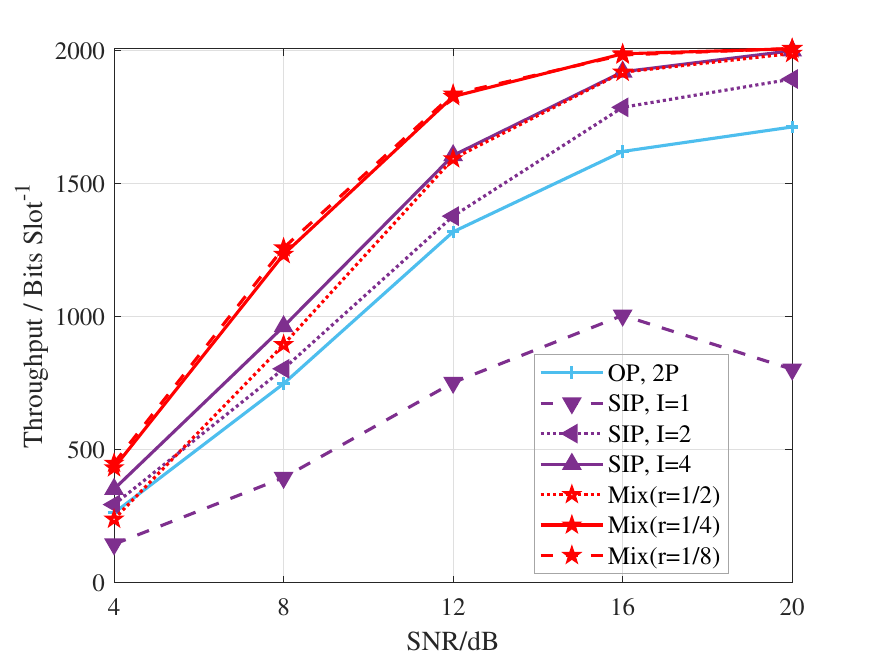}}
       \end{minipage}
       \caption{Throughput performance evaluations of our proposed neural receiver under $v=108$ km/h and $\text{ds}=363$ ns.}
       \vspace{-1em}
       \label{fig_tv_comp_tp}
     \end{figure}

The throughput performance of the proposed scheme is further illustrated in \figref{fig_tv_comp_tp}. Throughput is defined as the number of correctly received bits per slot, expressed by
\begin{equation}
       \label{equ_throughput}
       R = N_{\text{RB}} \times N_{\text{RE}} \times \Omega \times r \times Q \times \left( {1 - {\text{BLER}}} \right) ,
\end{equation}
where $N_{\text{RE}}=12\times14=168$ denotes the number of time-frequency REs per RB, and $\Omega$ represents the proportion of REs allocated to data symbols. Specifically, $\Omega=1$ is employed for both the SIP baseline and our proposed mix scheme, while $\Omega=12/14$ is used for the OP baseline. According to the figure, both the iterative SIP receiver with $I=4$ and our proposed receivers achieve approximately 15\% throughput improvement over the OP baseline at high SNRs, owing to the absence of dedicated pilot overhead. Furthermore, in the low-SNR regime, the proposed receivers with $r \in \{\frac{1}{4},\frac{1}{8}\}$ provide improved throughput relative to the SIP receiver, which aligns with the corresponding BLER performance gains.

\begin{table}[t]
       \caption{Complexity Analysis}
       \label{tab_complexity}
       \centering
       \begin{tabular}{lccc}
         \toprule
         \multirow{2}{*}{Scheme} & \multicolumn{2}{c}{Trainable Parameters} & Runtime \\
         & Channel estimation & Detection & (ms) \\
         \midrule
         Mix($r=1/4$) & $3.06 \times 10^5$ & \multirow{2}{*}{$1.5 \times 10^5$} & 5.67 \\
         Mix($r=1/8$) & $3.19 \times 10^5$ &  & 5.42 \\
         SIP, $I=4$ & -- & -- & 17.62 \\
         \bottomrule
       \end{tabular}
\end{table}

Moreover, a complexity analysis is conducted, as summarized in \tabref{tab_complexity}. As shown in the table, reducing the resource allocation ratio from $r=\frac{1}{4}$ to $r=\frac{1}{8}$ slightly increases the number of trainable parameters in the channel estimation network due to the inclusion of an additional interpolation module, while the preceding evaluations indicate broadly comparable performance between the two configurations. 

The SIP baseline employing $I=4$ iterations is also included for comparison. Note that the SIP receiver is evaluated under the assumption that channel statistics consistent with the testing scenario is always available, corresponding to a matched condition. Under this setting, the LMMSE-based channel estimation achieves performance comparable to that of the DL-based alternative, as demonstrated in \cite{SIPLi}. Therefore, the DL-enhanced refinement is not employed, and the SIP baseline contains no trainable parameters. Accordingly, the comparisons focus on processing latency. 
The results indicate that the proposed mix schemes require only a single round of channel estimation, detection, and decoding, resulting in an inference latency of less than one-third of that required by the iterative SIP receiver. 
Overall, the proposed scheme achieves an effective tradeoff between performance and computational efficiency under time-varying channel conditions.

\vspace{-1em}
\subsection{Generalization Capability}
\label{sec_simu_mm}
This subsection investigates the generalization capability of our proposed neural receiver, aiming to provide a comprehensive assessment of the feasibility and robustness of the developed transmission scheme under varying conditions. The OP receiver and the iterative SIP receivers with $I\in \{1,2,4\}$ are retained as baseline references. For the OP scheme, a ``1P'' pilot pattern as illustrated in \figref{fig_op_pattern}(b) is adopted to reflect the low-mobility scenarios considered in this subsection. In addition, the developed mix scheme employs a resource allocation ratio of $r=\frac{1}{4}$.

\begin{figure}[t]
       \centering
       \begin{minipage}{3in}
         \centerline{\includegraphics[width=3in]{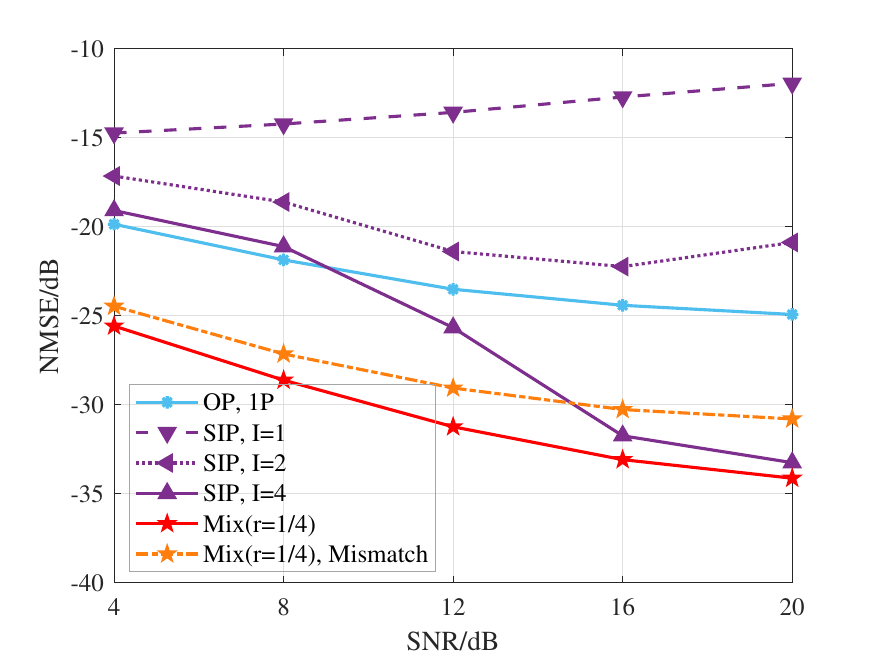}}
         \centerline{\small{(a) NMSE of channel estimation}}
       \end{minipage}
       \hfill
       \begin{minipage}{3in}
         \centerline{\includegraphics[width=3in]{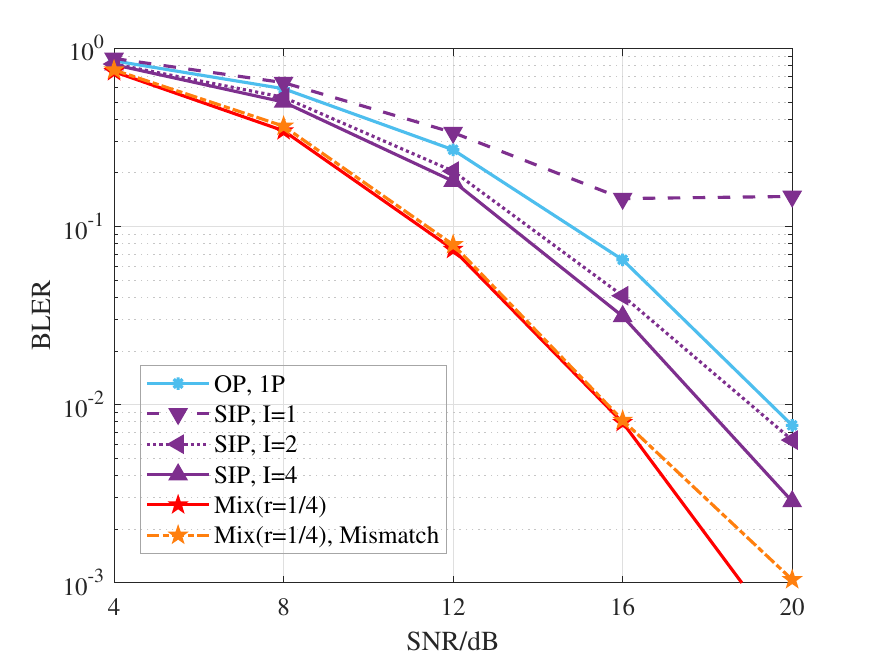}}
         \centerline{\small{(b) BLER of symbol demapping}}
       \end{minipage}
       \caption{Performance comparisons between our proposed neural receiver and the baseline receivers under $v=36$ km/h and $\text{ds}=93$ ns.}
       \label{fig_mm_low}
     \end{figure}

\subsubsection{Adaptability to Mismatched Channel Statistics}
The testing scenario is first configured with a UE velocity of $v=36$ km/h and a delay spread of $\text{ds}=93$ ns, where both matched and mismatched channel conditions are considered. Under the matched condition, the ``Mix(\emph{r}=1/4)'' model is trained using data collected under the current testing channel setting, and the matched channel correlation matrices are assumed to be available for the OP and SIP baselines. Under the mismatched condition, the neural receiver trained under the scenario described in \secref{sec_simu_tv}, i.e., $v=108$ km/h and $\text{ds}=363$ ns, is directly evaluated in the current environment, labeled as ``Mix(\emph{r}=1/4), Mismatch.''

As shown in \figref{fig_mm_low}(a), due to discrepancies in time- and frequency-domain channel variations between the training and testing scenarios, the mismatched neural receiver exhibits degraded channel estimation accuracy relative to the matched model. Nevertheless, the proposed mix scheme continues to offer a superior balance between performance and computational complexity compared with the SIP baseline. 
Moreover, \figref{fig_mm_low}(b) indicates that the NMSE degradation under mismatch has only a marginal impact on demapping performance. In particular, in the low-SNR regime, the mismatched model achieves BLER performance comparable to that of the matched counterpart. 
In addition, consistent with earlier observations, the mix schemes outperform the iterative SIP receivers in terms of BLER, highlighting the effectiveness of the proposed frame design in improving the reliability of soft information and mitigating error propagation during decoding. Overall, these results demonstrate that the proposed neural receiver exhibits strong robustness to mismatched channel statistics.

\begin{figure}[t]
       \centering
       \begin{minipage}{3in}
         \centerline{\includegraphics[width=3in]{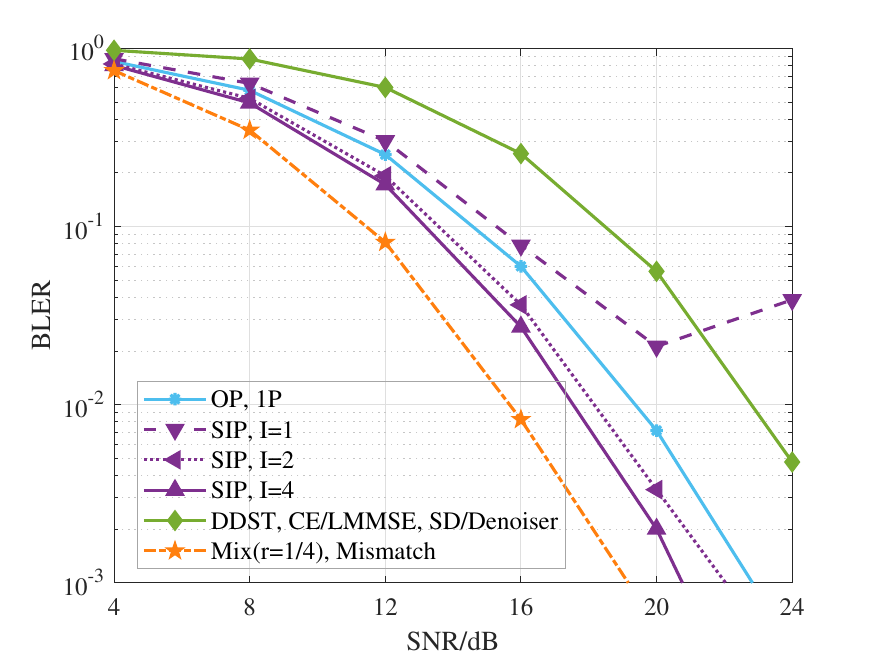}}
       \end{minipage}
       \caption{Performance comparisons between our proposed neural receiver and the enhanced DDST receiver when $\rho=0.3$ is set under $v=0$ km/h and $\text{ds}=93$ ns.}
       \label{fig_mm_block}
     \end{figure}

\subsubsection{Applicability to Block-fading Scenarios}
To further evaluate robustness, a more extreme mismatched condition is considered, in which the neural receiver trained under $v=108$ km/h and $\text{ds}=363$ ns is directly deployed in a block-fading environment characterized by $v=0$ km/h and $\text{ds}=93$ ns. The enhanced DDST receiver proposed in \secref{sec_block_enhance} is also included for comparison. This scheme can be interpreted as a special case of the mix transmission framework with $r=1$,\footnote{Note that the power allocation for the developed DDST receiver was previously set to $\rho=1/7$ in \secref{sec_simu_block} to demonstrate comparable channel estimation performance between the DDST mechanism and the OP scheme with TDM. In contrast, the current evaluations adopt a unified empirical setting of $\rho=0.3$ to ensure a fair comparison across different DDST-based schemes.} except that the LMMSE-based channel estimation is adopted given the quasi-static assumption. As shown in \figref{fig_mm_block}, the BLER results indicate that the mismatched neural receiver outperforms both reference baselines and the $r=1$ case. This observation highlights the effectiveness of the proposed frame design and the developed ViT architecture in mitigating coupled interference and sustaining robust receiver performance.

\section{Conclusion}
\label{sec_conclusion}
This paper presented a low-overhead transmission framework to address the excessive pilot overhead and computational burden associated with the OP and SIP schemes. An enhanced DDST receiver, comprising the LMMSE-based channel estimation and the denoiser-based detection, is first developed to improve performance under the quasi-static assumption. To enable feasibility in practical wireless environments, a mix transmission scheme is further designed, based on which a neural receiver is proposed. The receiver leverages the self-attention mechanism to effectively exploit the inherent orthogonal structure of high-dimensional observations. Simulation results demonstrate that the proposed neural receiver under the mix transmission scheme successfully combines the pilot overhead reduction of superimposed transmission with the interference mitigation capability of data-dependent structures, thereby achieving significant performance gains and a favorable performance-complexity balance over OP- and SIP-based receivers across diverse time-varying scenarios.

\ifCLASSOPTIONcaptionsoff
  \newpage
\fi


\begin{thebibliography}{99}

  \bibitem{OPMar}
  T.~L.~Marzetta, ``{Noncooperative cellular wireless with unlimited numbers of base station antennas},'' \emph{IEEE Trans. Wireless Commun.}, vol.~9, no.~11, pp. 3590--3600, Nov. 2010.

  \bibitem{GenZhou}
  X.~Zhou, L.~Liang, J.~Zhang, P.~Jiang, Y.~Li, and S.~Jin, ``{Generative diffusion models for high dimensional channel estimation},'' \emph{IEEE Trans. Wireless Commun.}, vol.~24, no.~7, pp. 5840--5854, Jul. 2025.

  \bibitem{OPTru}
  S.~Trushkov, V.~Kuptsov, O.~Shmonin, K.~Ponur, G.~Serebryakov, and A.~Blagodarnyi, ``{Pilot overhead reduction for antenna ports in MIMO OFDM systems using high-resolution map},'' in \emph{Proc. IEEE Int. Black Sea Conf. Commun. Netw. (BlackSeaCom)}, Tbilisi, Georgia, Jun. 2024, pp. 66--71.

  \bibitem{NRCite}
  {3GPP}, ``{NR; Physical channels and modulation},'' 3GPP TS 38.211, Tech. Rep., 2023.
  
  \bibitem{SIPXie}
  M.~Xie, X.~Yu, K.~Wang, J.~Zhang, X.~Dang, and C.~Yuen, ``{Superimposed pilots for cell-free massive MIMO over spatial-correlated Rician fading channels},'' \emph{IEEE Trans. Wireless Commun.}, vol.~23, no.~12, pp. 19\,537--19\,552, Dec. 2024.
  
  \bibitem{SIPAit}
  F.~Ait~Aoudia and J.~Hoydis, ``{End-to-end learning for OFDM: From neural receivers to pilotless communication},'' \emph{IEEE Trans. Wireless Commun.}, vol.~21, no.~2, pp. 1049--1063, Feb. 2022.
  
  \bibitem{SIPXiao}
  H.~Xiao \emph{et al.}, ``{Interference cancellation based neural receiver for superimposed pilot in multi-layer transmission},'' \emph{China Commun.}, vol.~22, no.~1, pp. 75--88, Jan. 2025.

  \bibitem{GenZhou2}
  X.~Zhou \emph{et al.}, ``{Conditional diffusion model-enabled scenario-specific neural receivers for superimposed pilot schemes},'' \emph{arXiv preprint arXiv:2511.01173}, 2025.

  \bibitem{SIPMa}
  J.~Ma, C.~Liang, C.~Xu, and L.~Ping, ``{On orthogonal and superimposed pilot schemes in massive MIMO NOMA systems},'' \emph{IEEE J. Sel. Areas Commun.}, vol.~35, no.~12, pp. 2696--2707, Dec. 2017.

  \bibitem{SIPJing}
  X.~Jing, M.~Li, H.~Liu, S.~Li, and G.~Pan, ``{Superimposed pilot optimization design and channel estimation for multiuser massive MIMO systems},'' \emph{IEEE Trans. Veh. Technol.}, vol.~67, no.~12, pp. 11\,818--11\,832, Dec. 2018.

  \bibitem{SIPQian}
  C.~Qian, R.~Gu, W.~Xu, J.~Xu, and X.~You, ``{Enhancing wideband multiuser MIMO uplink using superimposed pilots: Joint receiver design},'' \emph{IEEE Wireless Commun. Lett.}, vol.~13, no.~4, pp. 1138--1142, Apr. 2024.

  \bibitem{SIPLi}
  X.~Li \emph{et al.}, ``{Learning-aided iterative receiver for superimposed pilots: Design and experimental evaluation},'' \emph{IEEE Trans. Wireless Commun.}, vol.~25, pp. 13\,864--13\,880, 2026.

  \bibitem{SIPLi2}
  X.~Li, X.~Zhou, J.~Zhang, C.-K. Wen, and S.~Jin, ``{AI-driven iterative receiver for superimposed pilot schemes in MIMO-OFDM systems},'' in \emph{Proc. IEEE Wireless Commun. Netw. Conf. (WCNC)}, Milan, Italy, Mar. 2025, pp. 1--6.

  \bibitem{GenZhang}
  R.~Zhang \emph{et al.}, ``{Score-based conditional flow models for MIMO receiver design with superimposed pilots},'' \emph{IEEE Open J. Commun. Soc.}, vol.~7, pp. 3331--3345, 2026.

  \bibitem{DDSTGho2}
  M.~Ghogho, D.~McLernon, E.~Alameda-Hernandez, and A.~Swami, ``{Channel estimation and symbol detection for block transmission using data-dependent superimposed training},'' \emph{IEEE Signal Process. Lett.}, vol.~12, no.~3, pp. 226--229, Mar. 2005.

  \bibitem{DDSTWu}
  Y.~Wu and S.~Sugiura, ``{Reduced-overhead channel estimation and iterative detection of FTN signaling based on pilot superimposition and spectral interference alignment},'' in \emph{Proc. IEEE Global Commun. Conf. (GLOBECOM)}, Taipei, Taiwan, Dec. 2025, pp. 5820--5825.

  \bibitem{DDSTTvGho}
  M.~Ghogho and A.~Swami, ``{Estimation of doubly-selective channels in block transmissions using data-dependent superimposed training},'' in \emph{Proc. 14th Eur. Signal Process. Conf. (EUSIPCO)}, Florence, Italy, Sept. 2006, pp. 1--5.

  \bibitem{DDSTTvHe}
  S.~He and J.~K.~Tugnait, ``{On doubly selective channel estimation using sperimposed training and discrete prolate spheroidal sequences},'' \emph{IEEE Trans. Signal Process.}, vol.~56, no.~7, pp. 3214--3228, Jul. 2008.

  \bibitem{DDSTTvCar}
  R.~Carrasco-Alvarez, R.~Parra-Michel, and A.~G.~Orozco-Lugo, ``{Enhanced time-varying channel estimation based on two dimensional basis projection and self-interference suppression},'' in \emph{Proc. IEEE 11th Int. Workshop Signal Process. Adv. Wireless Commun. (SPAWC)}, Marrakech, Morocco, Jun. 2010, pp. 1--5.

  \bibitem{DDSTChan}
  K.-C.~Chan, W.-C.~Huang, C.-P.~Li, and H.-J.~Li, ``{Investigation on data identification problem for data-dependent superimposed training},'' in \emph{Proc. IEEE 75th Veh. Technol. Conf. (VTC Spring)}, Yokohama, Japan, May 2012, pp. 1--5.

  \bibitem{DDSTDou}
  G.~Dou, C.~Li, J.~Gao, and F.~Guo, ``{Constellation rotation and symbol detection for data-dependent superimposed training},'' \emph{Electron. Lett.}, vol.~50, no.~25, pp. 1939--1940, Dec. 2014.

  \bibitem{DDSTQing}
  C.~Qing, L.~Dong, L.~Wang, J.~Wang, and C.~Huang, ``{Joint model and data-driven receiver design for data-dependent superimposed training scheme with imperfect hardware},'' \emph{IEEE Trans. Wireless Commun.}, vol.~21, no.~6, pp. 3779--3791, Jun. 2022.

  \bibitem{DDSTGho}
  M.~Ghogho and A.~Swami, ``{Channel estimation for MIMO systems using data-dependent superimposed training},'' in \emph{Proc. 42nd Annu. Allerton Conf. Commun., Control, Comput. (Allerton)}, Monticello, IL, USA, Sept. 2004, pp. 70--79.

  \bibitem{DDSTKam}
  A.~Kammoun, K.~Abed-Meraim, and S.~Affes, ``{Performance of linear receivers based on superimposed training},'' in \emph{Proc. IEEE 8th Int. Workshop Signal Process. Adv. Wireless Commun. (SPAWC)}, Helsinki, Finland, Jun. 2007, pp. 1--5.

  \bibitem{SIPZou}
  J.~Zou, J.~Xiao, Q.~Mao, S.~Liu, B.~Xiao, and Y.~Liang, ``{Deep receiver for multi-layer data transmission with superimposed pilots},'' in \emph{Proc. IEEE Int. Conf. Acoust., Speech Signal Process. (ICASSP)}, Hyderabad, India, Apr. 2025, pp. 1--5.

  \bibitem{WHKay}
  S.~M. Kay, \emph{{Fundamentals of statistical signal processing}}.\hskip 1em plus 0.5em minus 0.4em\relax Upper Saddle River, NJ, USA: Prentice-Hall, 1993.

  \bibitem{BlockSDHon}
  M.~Honkala, D.~Korpi, and J.~M.~J.~Huttunen, ``{DeepRx: Fully convolutional deep learning receiver},'' \emph{IEEE Trans. Wireless Commun.}, vol.~20, no.~6, pp. 3925--3940, Jun. 2021.

  \bibitem{BlockSDGui}
  G.~Gui, H.~Huang, Y.~Song, and H.~Sari, ``{Deep learning for an effective nonorthogonal multiple access scheme},'' \emph{IEEE Trans. Veh. Technol.}, vol.~67, no.~9, pp. 8440--8450, Sept. 2018.

  \bibitem{BlockSDEmi}
  A.~Emir, F.~Kara, H.~Kaya, and H.~Yanikomeroglu, ``{DeepMuD: Multi-user detection for uplink grant-free NOMA IoT networks via deep learning},'' \emph{IEEE Wireless Commun. Lett.}, vol.~10, no.~5, pp. 1133--1137, May 2021.
  
  \bibitem{TvCEDos}
  A.~Dosovitskiy \emph{et al.}, ``{An image is worth 16x16 words: Transformers for image recognition at scale},'' \emph{arXiv preprint arXiv:2203.11854}, 2020.

  \bibitem{TvCELiu}
  F.~Liu, J.~Zhang, P.~Jiang, C.-K.~Wen, and S. Jin, ``{CE-ViT: A robust channel estimator based on vision transformer for OFDM systems},'' in \emph{Proc. IEEE Global Commun. Conf. (GLOBECOM)}, Kuala Lumpur, Malaysia, Dec. 2023, pp. 4798--4803.

  \bibitem{TvCEVas}
  A.~Vaswani \emph{et al.}, ``{Attention is all you need},'' in \emph{Proc. Adv. Neural Inf. Process. Syst. (NIPS)}, Long Beach, CA, USA, Jun. 2017, pp. 5998--6008.

  \bibitem{TvCEYuan}
  K.~Yuan \emph{et al.}, ``{Incorporating convolution designs into visual transformers},'' in \emph{Proc. IEEE/CVF Int. Conf. Comput. Vis. (ICCV)}, Oct. 2021, pp. 579--588.

  \bibitem{TvCELi}
  Y.~Li \emph{et al.}, ``{LocalViT: Analyzing locality in vision transformers},'' in \emph{Proc. IEEE/RSJ Int. Conf. Intell. Robots Syst. (IROS)}, Detroit, MI, USA, Oct. 2023, pp. 9598--9605.

  \bibitem{TvCELuan}
  D.~Luan and J.~Thompson, ``{Attention based neural networks for wireless channel estimation},'' in \emph{Proc. IEEE 95th Veh. Technol. Conf. (VTC2022-Spring)}, Helsinki, Finland, Jun. 2022, pp. 1--5.

  \bibitem{TvCEZhao}
  M.~Zhao, G.~Cao, X.~Huang, and L.~Yang, ``{Hybrid transformer-CNN for real image denoising},'' \emph{IEEE Signal Process. Lett.}, vol.~29, pp. 1252--1256, 2022.

  \bibitem{CDLCite}
  {3GPP}, ``{Study on channel model for frequencies from 0.5 to 100 GHz},'' 3GPP TR 38.901, Tech. Rep., 2022.

  \bibitem{SionnaHoy}
  J.~Hoydis \emph{et al.}, ``{Sionna: An open-source library for next-generation physical layer research},'' \emph{arXiv preprint arXiv:2203.11854}, 2022.

  \bibitem{MMSEICStu}
  C.~Studer, S.~Fateh, and D.~Seethaler, ``{ASIC implementation of soft-input soft-output MIMO detection using MMSE parallel interference cancellation},'' \emph{IEEE J. Solid-State Circuits}, vol.~46, no.~7, pp. 1754--1765, Jul. 2011.

  \bibitem{LMMSECite}
  Y.~Liu, Z.~Tan, H.~Hu, L.~J. Cimini, and G.~Y. Li, ``{Channel estimation for OFDM},'' \emph{IEEE Commun. Surveys Tut.}, vol.~16, no.~4, pp. 1891--1908, May 2014.

  
\end{thebibliography}

\end{document}